\documentclass[twocolumn,showpacs,preprintnumbers,prd,nofootinbib]{revtex4-1}
\usepackage{graphicx}	
\usepackage{dcolumn}	
\usepackage{bm}			
\usepackage{hyperref}
\usepackage{amsmath}
\usepackage{verbatim}
\usepackage{slashed}
\usepackage{mathtools}
\usepackage{tabularx}

\hypersetup{
    colorlinks,
    citecolor=blue,
    filecolor=black,
    linkcolor=blue,
    urlcolor=black
}
\usepackage{slashed}
\usepackage{cancel}

\begin{document}

\author{Yi Chung}
\email{yi.chung@mpi-hd.mpg.de}
\affiliation{Max-Planck-Institut f\"ur Kernphysik, Saupfercheckweg 1, 69117 Heidelberg, Germany}

\title{Two coincidences are a clue: Probing a GeV-scale dark QCD sector}

\begin{abstract}
The similarity between the dark matter and baryon energy densities suggests an existence of a dark sector analogous to QCD. In addition, small-scale structure anomalies can be addressed by dark matter self-interactions with cross sections comparable to those of QCD. Both observations point toward a GeV-scale dark QCD sector. Motivated by these two coincidences, we investigate the parameter space of a distinctive chiral dark QCD model featuring a MeV-scale dark photon with axial-vector couplings. We also discuss a possible third coincidence associated with the latest measurement of $N_{\rm eff}$. Current constraints leave a finite and testable region of parameter space that can be probed by future experiments such as the Gamma Factory.
\end{abstract}

\maketitle

\section{Introduction}

The nature of dark matter remains the most intriguing problems of physics beyond the Standard Model (SM). To date, all observational evidence comes from its gravitational effects on cosmology and astronomy. Although the available information is limited, current measurements reveal two striking similarities that may provide important clues to uncovering the mysterious dark sector.


First, the measurement of the cosmic microwave background (CMB) provides the most precise quantitative measurement of dark matter, namely its energy density, $\Omega_c \approx 0.26$ \cite{Planck:2018vyg}. While this value alone can be accommodated by a wide variety of dark matter models, a comparison with the observed baryon energy density, $\Omega_b \approx 0.05$, reveals a surprising ratio $\Omega_c / \Omega_b \approx 5$. This coincidence, known as the dark matter–baryon coincidence problem, hints at a deep connection between the dark sector and the QCD sector.

Previous attempts to address this problem include mirror dark sectors \cite{Hodges:1993yb,Berezhiani:1995am,Foot:2003jt,An:2009vq,Farina:2015uea,GarciaGarcia:2015pnn,Lonsdale:2018xwd,Ibe:2019ena,Bodas:2024idn} and dark grand unification \cite{Murgui:2021eqf,Chung:2024nnj}, which can simultaneously realize comparable number densities, $n_D \approx n_B$, and masses, $m_D \approx m_B$, of dark matter and baryons, thereby yielding similar relic abundances. Other mechanisms, such as infrared fixed points \cite{Bai:2013xga,Newstead:2014jva,Ritter:2022opo,Ritter:2024sqv} and QCD-induced dark matter masses \cite{Chung:2024ezq}, can also introduce GeV-scale dark matter candidates. When embedded within frameworks of Asymmetric Dark Matter (ADM) \cite{Spergel:1984re,Nussinov:1985xr,Gelmini:1986zz,Barr:1990ca,Barr:1991qn,Kaplan:1991ah,Gudnason:2006ug,Gudnason:2006yj,Kaplan:2009ag,Davoudiasl:2012uw,Petraki:2013wwa,Zurek:2013wia}, these mechanisms can further explain the observed coincidence.\footnote{The coincidence can also be explained without requiring comparable number densities and masses—for example, if the production mechanism and mass are correlated \cite{Rosa:2022sym}, or if novel dynamics arise to balance the two energy densities \cite{Brzeminski:2023wza,Banerjee:2024xhn}.}
A generic feature of these scenarios is asymmetric dark baryon dark matter originating from a GeV-scale dark QCD sector.

However, one coincidence may simply be accidental. An additional hint toward a GeV-scale dark sector comes from astronomy, the long-standing small-scale structure tensions known as the core–cusp problem. While the discrepancy between observations \cite{Flores:1994gz,Moore:1994yx,Moore:1999gc,Walker:2011zu,Tollerud:2014zha,deBlok:2001hbg,deBlok:2002vgq,Simon:2004sr,Gentile:2004tb,Newman:2012nv,Newman:2012nw} and numerical simulations \cite{Dubinski:1991bm,Navarro:1995iw,Navarro:1996gj} remains inconclusive due to the complexity of baryonic feedback processes \cite{Martizzi:2011aa,Governato:2012fa,Benitez-Llambay:2019wfi}, its solution—should it persist—could provide valuable insight into the microscopic nature of dark matter.

Among the proposed solutions, self-interacting dark matter (SIDM) remains an attractive possibility \cite{Spergel:1999mh,Dave:2000ar}, supported by numerical simulations \cite{Vogelsberger:2012ku,Rocha:2012jg,Peter:2012jh,Elbert:2014bma}. To alleviate the core–cusp problem, SIDM requires a self-interaction cross section $\sigma_D/m_D \sim 1~\mathrm{cm}^2/\mathrm{g}$, which is intriguingly comparable to the nucleon self-scattering cross section, $\sigma_B/m_B \sim 12~\mathrm{cm}^2/\mathrm{g}$. From naive dimensional analysis, both cross sections point toward the similar underlying mass scale of $\mathcal{O}(1)$ GeV. We refer to this resemblance as the second coincidence.

However, small discrepancies observed at cluster scales \cite{Newman:2012nv,Newman:2012nw} point to a nontrivial velocity dependence. It has been shown that such a velocity dependence can be realized with a light dark photon of mass $\mathcal{O}(10)$ MeV \cite{Kaplinghat:2015aga}, motivating a modification beyond SM QCD sector. With this motivation, we consider a dark QCD sector with a dark photon of a new $U(1)_D$ gauge group, which serves both as the mediator of the dark force and as a portal to the visible sector. Moreover, motivated by the observation that the required dark photon mass is $\sim g f_\pi$ with a typical $g\sim\mathcal{O}(0.1)$, where $f_\pi=92$ MeV is the pion decay constant, representing the symmetry breaking scale of QCD, we consider a chiral dark QCD sector where the $U(1)_D$ gauge group is dynamically broken by dark QCD, featuring a single characteristic scale.

With these two observational coincidences
\begin{align}\label{twoco}
\text{Energy density :}&&n_D\,m_D&\approx 5 \times n_B\,m_B~,\\
\text{Self-interaction :}&&\sigma_D/m_D &\approx 0.1\times\sigma_B/m_B~,
\end{align}
both pointing toward a GeV-scale dark QCD sector, we aim to investigate whether a common parameter space can accommodate them simultaneously and remain phenomenologically viable. We begin with a simplified model of a chiral dark QCD sector motivated by these coincidences. We then explore the parameter space that yields the required dark matter masses and velocity-dependent self-interaction cross sections. Finally, we discuss constraints on dark baryons and dark photons, with particular emphasis on $N_{\rm eff}$, and identify promising future experiments capable of testing this scenario.

\section{A simplified model of dark sector}

Solutions to the dark matter–baryon energy density coincidence typically require a nontrivial dark QCD sector. In this work, we do not attempt to construct a complete solution to this problem. Instead, we refer the reader to Refs.~\cite{Murgui:2021eqf,Chung:2024nnj,Bai:2013xga,Newstead:2014jva,Ritter:2022opo,Ritter:2024sqv,Chung:2024ezq} for model-building efforts that go beyond simple mirror dark sector setups. These studies feature a wide variety of dark QCD sectors that differ substantially from the SM QCD sector. Such structures not only motivate the chiral dark QCD sector considered here, but also are likely required to account for the distinct properties of dark matter and baryons despite their comparable energy densities.

In this study, we therefore seek a dark QCD model capable of simultaneously accommodating the requirements implied by the two coincidences. While explaining the energy density coincidence typically calls for a GeV-scale dark baryon arising from a strongly coupled dark sector, the self-interaction requirement points to GeV-scale dark matter accompanied by a MeV-scale dark photon instead. Rather than introducing the dark photon mass by hand, we assume that it originates from the same dynamical source as the dark matter mass, following the general spirit of chiral dark QCD models \cite{Harigaya:2016rwr,Co:2016akw,Ibe:2021gil}, in which the dark $U(1)_D$ gauge symmetry is dynamically broken by the dark quark condensate and the dark sector is characterized by a single dynamical scale $f$. However, we adopt a different structural realization motivated by Ref.~\cite{Chung:2024ezq}, as discussed in the Appendix.

We consider a dark gauge group $SU(N)_D \times U(1)_D$ with two dark Weyl fermions, $\psi_L$ and $\bar{\psi}_R$, transforming as $(\mathbf{N},1/2)$ and $(\overline{\mathbf{N}},1/2)$, respectively.\footnote{Note that this simplified model is anomalous, and we expect additional states to be present in a complete chiral dark QCD model. These states, however, play irrelevant role in the phenomenology considered here, as will be justified in the Appendix.} Once the dark color gauge group $SU(N)_D$ becomes strongly coupled at the GeV scale, the fermions $\psi_L$ and $\bar{\psi}_R$ form a condensate. The dynamics of the dark quark condensate $\psi_L\bar{\psi}_R$ can be conveniently described by introducing an auxiliary complex scalar field $\Phi$, transforming as $(\mathbf{1},1)$. The formalism follows the Nambu–Jona-Lasinio (NJL) model \cite{Nambu:1961tp,Nambu:1961fr}, which provides a useful effective description for strongly coupled theory such as QCD \cite{Klevansky:1992qe,Hatsuda:1994pi}. The effective Lagrangian of the simplified model can then be written as
\begin{align}
\mathcal{L}_{\rm DS}=&-\frac{1}{4}F'_{\mu\nu}F'^{\mu\nu}+
|D_\mu \Phi|^2+\mu^2|\Phi|^2-\lambda|\Phi|^4\nonumber\\
&+i\bar{\psi}_{L,R}\,\slashed{D}\,\psi_{L,R}+
Y_\psi\,\bar{\psi}_L\,\Phi\, \psi_R+\textrm{h.c.}~,
\end{align}
where $Y_\psi\sim 4\pi/\sqrt{N}$ is the Yukawa coupling between the dark quarks and their condensate, as estimated within the NJL framework.

The condensate $\Phi$ acquires a vacuum expectation value (VEV) $\langle\Phi\rangle =f$ from the strong dynamics, generating masses for both the dark quark and the dark photon:
\begin{align}
m_\psi=Y_\psi\,f\,,\quad m_{\gamma'}=g_D\,f~
\end{align}
with $g_D$ denoting the gauge coupling of the $U(1)_D$. As the theory is confining, the degrees of freedom below the confinement scale should be baryons rather than quarks. The dark matter candidate is therefore the dark baryon $D$, a spin-$N/2$ state composed of $N$ dark quarks \cite{Antipin:2015xia,Morrison:2020yeg,Garani:2021zrr}, with its mass given by
\begin{align}
m_D\equiv Y_D\,f\,\sim N\,Y_\psi\,f\,,\quad \text{where  $Y_D\sim\mathcal{O}(4\pi)$ .}
\end{align}
We take the number of colors $N = 3$ as our benchmark, leaving 3 independent parameters for dark QCD sector: (1) \textbf{dark QCD VEV $\bm{f}$}, (2) \textbf{dark matter mass $\bm{m_D}$}, (3) \textbf{dark photon mass $\bm{m_{\gamma'}}$}. The couplings are then determined by $Y_D(g_D) = m_D(m_{\gamma'})/f$.

Besides featuring a single dynamical scale, another important property of our chiral dark sector is that the dark photon couples exclusively to the axial-vector current,
\begin{align}
\mathcal{L}_{\gamma'}= i\,\frac{1}{2}\,g_D A'_\mu \bar{\psi}\gamma^\mu\gamma^5\psi~,
\end{align}
which differs from conventional dark photon models with vector couplings. The dark photon coupling to the dark baryon $D$ is, in general, nontrivial and involves multiple form factors due to its spin-$N/2$ nature, potentially leading to distinct phenomenology that requires dedicated study. However, for the processes considered in this work, the transfer momentum is small ($q^2 \ll m_D^2$), such that the leading gauge interaction is determined solely by the total charge $Q_D = N/2$, with the same kinetic structure as a single dark quark. As will be discussed below, this purely axial-vector coupling plays an important role in dark matter direct detection.

In addition, we need a portal between dark sector and the visible sector. In this study, we consider the kinetic mixing between the dark photon and the SM photon \cite{Holdom:1985ag}
\begin{align}
\mathcal{L}_{\rm portal}=-\frac{\epsilon}{2}F'_{\mu\nu}F^{\mu\nu}\implies \epsilon e A'_\mu J^\mu_{\rm QED}~,
\end{align}
which introduces interactions between the dark sector and the SM. The \textbf{kinetic mixing parameter $\bm{\epsilon}$} is the fourth and final parameter of this simplified model and plays a crucial role in searches for the dark sector.

\begin{figure*}[t]
  \centering
  \includegraphics[width=\textwidth]{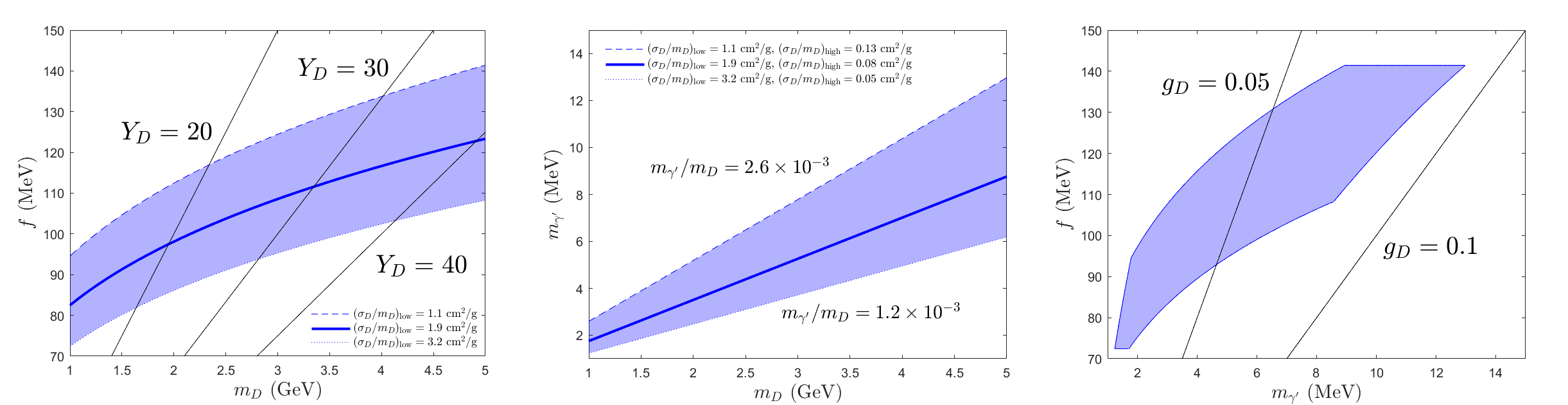}
  \caption{The favored parameter space (shaded band) for the chiral dark QCD model to explain the two coincidences. Left: On the $(m_D,f)$ plane, the blue band is determined by the $(\sigma_D/m_D)_{\rm low}$ with values shown in the legend. The three black lines represent different values of Yukawa coupling $Y_D$ of the dark matter. Center: The ratio between $(\sigma_D/m_D)_{\rm high}$ and $(\sigma_D/m_D)_{\rm low}$ determine the required value of $m_{\gamma'}/m_D$ as shown on the $(m_{\gamma'},m_D)$ plane. Right: The favored parameter space on the $(m_{\gamma'},f)$ plane with black lines representing different values of gauge coupling $g_D$.\label{fig:versus}}
\end{figure*}

\section{Inputs from the two coincidences}

Having established a simplified model of dark baryons and dark photons motivated by the two coincidences, we now proceed to analyze the region of parameter space $(f, m_D, m_{\gamma'}, \epsilon)$ favored by these considerations.

\subsection{Energy density coincidence}

We begin with the energy densities coincidence. Solutions to this coincidence typically require both comparable number densities, $n_D \approx n_B$, and masses, $m_D \approx m_B$. The first condition can be naturally addressed within the framework of ADM and its associated cosmological evolution. As we will discuss later, this requirement has important implications for the mixing parameter $\epsilon$.

The second condition requires the ratio $m_D/m_B$ to be of $\mathcal{O}(1)$. For our parameter space analysis, we restrict this ratio to be larger than unity but not exceed the observed energy density ratio, $\Omega_c / \Omega_b \approx 5$, in order to avoid introducing unnatural fine-tuning among parameters. The viable dark matter mass is thus confined to a narrow range, $m_D = 1$–$5~\mathrm{GeV}$, in this analysis.

\subsection{Dark matter self-interaction cross section}

Next, we look into the second coincidence from astronomy. If DM self-interactions are responsible for resolving the core-cusp problem, $\sigma_D/m_D \sim 0.1-1$ cm$^2/$g is typically required. This requirement can be further refined by considering two distinct velocity regimes.

In the low-velocity regime ($v \sim 30$–$200\ \text{km/s}$), relevant for dwarf and low surface brightness galaxies, the best-fit value is $(\sigma_D/m_D)_{\rm low}= 1.9^{+0.6}_{-0.4}$ cm$^2/$g \cite{Kaplinghat:2015aga}. In contrast, observations at cluster scales, corresponding to the high-velocity regime ($v=1458$ km/s), prefer a smaller cross section, $(\sigma_D/m_D)_{\rm high}= 0.082^{+0.027}_{-0.021}$ cm$^2/$g \cite{Andrade:2020lqq}.

To analyze the cross section mediated by dark photon exchange, we consider the Born approximation with the transfer cross section given by \cite{Feng:2009hw}
\begin{align}
\sigma_D\approx \sigma_T^{\rm Born}=\frac{1}{2\pi}\frac{Q_D^4 g_D^4}{m_D^2v^4}\left({\rm ln}(1+R^2)-\frac{R^2}{1+R^2}\right),
\end{align}
where $R\equiv {m_Dv}/m_{\gamma'}$. The approximation is valid for $\alpha_Dm_D m_{\gamma'}\ll 1$, which will be justified later. We can then express it in terms of model parameters
\begin{align}
\frac{\sigma_D}{m_D}\approx\frac{Q_D^4}{4\pi}\frac{m_D}{f^4}\left[\frac{2}{R^4}\left({\rm ln}(1+R^2)-\frac{R^2}{1+R^2}\right)\right].
\end{align}
The function enclosed in brackets, denoted as $F(R)$, is $\approx 1$ in the limit $R\propto v\ll 1$. Therefore, the prefactor represents the cross section at low velocities. The function $F(R)$ drops with growing $R$ and suppresses the cross section at high velocities, explaining the required velocity-dependence. Such expression allow us to analyze the low- and high-velocity regimes separately.

In the low-velocity regime, where $F(R)=[\cdots]\approx 1$, the $(\sigma_D/m_D)_{\rm low}$ is simply a function of $m_D$ and $f$. Therefore, we can obtain (with $Q_D=3/2$)
\begin{align}
f=82.5~{\rm MeV}\left(\frac{1.9~\text{cm$^2/$g}}{(\sigma_D/m_D)_{\rm low}}\right)^\frac{1}{4}\left(\frac{m_D}{1~{\rm GeV}}\right)^\frac{1}{4}~.
\end{align}
This expression allows us to determine the VEV $f$ as a function of $m_D$, assuming the best-fit value of the low-velocity cross section from \cite{Kaplinghat:2015aga}. The shaded region corresponding to the $\pm 2\sigma$ range, $(\sigma_D/m_D)_{\rm low} = [1.1,3.2]$ cm$^2/$g, is shown in Fig.~\ref{fig:versus} (left).

The preferred scale of dark sector $f\sim 100$ MeV, similar to the QCD scale, is the direct consequence of the second coincidence. Also shown is the corresponding Yukawa coupling $Y_D$ along the favored parameter space. The resulting $Y_D\sim\mathcal{O}(4\pi)$ is consistent with expectations from strong dynamics, which further justifies the requirement of a strongly coupled dark sector.

Next, we turn to the high-velocity regime. With a fixed $(\sigma_D/m_D)_{\rm low}$, the value of $(\sigma_D/m_D)_{\rm high}$ is determined by the velocity-dependent suppression encoded in the function $F(R)$, where $R \equiv m_D v / m_{\gamma'}$. This time, it is a function of $m_D$ and $m_{\gamma'}$, independent of the VEV $f$.

We can then derive the relation between $m_D$ and $m_{\gamma'}$ using the ratio
\begin{align}
r\equiv\frac{(\sigma_D/m_D)_{\rm high}}{(\sigma_D/m_D)_{\rm low}}=\frac{F(R_h)}{F(R_\ell)}\approx F(R_h)
\end{align}
where $R_h\equiv {m_Dv_h}/m_{\gamma'}$ with $v_h$ = $1458$ km/s. The dark photon mass can then be expressed as function of dark matter mass as
\begin{align}
m_{\gamma'}=\frac{v_h}{R_h}m_D=4.9~{\rm MeV}\left(\frac{1}{F^{-1}(r)}\right)\left(\frac{m_D}{1~{\rm GeV}}\right)~.
\end{align}
For example, using the best-fit values of the cross sections, we find $F^{-1}(0.082 / 1.9) = 2.77$. The resulting parameter space is shown in Fig. \ref{fig:versus} (center), yielding the range $1.2 \times 10^{-3} < m_{\gamma'}/m_D < 2.6 \times 10^{-3}$. Given the preferred dark matter mass $m_D = 1$–$5$ GeV from the first coincidence,  a dark photon mass $m_{\gamma'}=1-13$ MeV is required to explain the small-scale structure.

Also shown in Fig. \ref{fig:versus} (right), is the plot of $f$ v.s. $m_{\gamma'}$, which gives us a glance on the gauge coupling $g_D$. The viable parameter space favors a weak coupling $g_D< 0.1$, justifying the usage of Born approximation.

\section{$N_{\rm eff}$ as the third coincidence?}

So far, we have derived a GeV-scale chiral dark QCD sector whose parameter space is motivated by the two coincidences. As the analysis so far relies only on dark sector properties inferred from gravitational effects, the results remain independent of the SM.

However, since the energy density coincidence favors an asymmetric dark matter scenario \cite{Spergel:1984re,Nussinov:1985xr,Gelmini:1986zz,Barr:1990ca,Barr:1991qn,Kaplan:1991ah,Gudnason:2006ug,Gudnason:2006yj,Kaplan:2009ag,Davoudiasl:2012uw,Petraki:2013wwa,Zurek:2013wia}, interactions between the dark sector and visible sectors in the early Universe are required to generate comparable number densities of dark baryons and QCD baryons. The detailed mechanism for asymmetry generation depends on the UV-complete model and is beyond the scope of this work. Nevertheless, a sizable entropy associated with the symmetric component of the dark sector is generically expected, which gives rise to additional constraints that we will address in this section. In our setup, this entropy is ultimately transferred to the lightest state in the dark sector, the MeV-scale dark photon $\gamma'$.

To prevent the Universe from being overclosed by the relic abundance of dark photons (assuming a sufficiently high reheating temperature), a nonzero kinetic mixing parameter $\epsilon$ is required, allowing dark photons to decay into SM particles. This decay has significant implications for the thermal history around $T \sim m_{\gamma'} = \mathcal{O}(\mathrm{MeV})$. Since MeV-scale dark photons decay predominantly into the electron–photon plasma, if the decay occurs after neutrino decoupling, it increases the ratio $T_\gamma/T_\nu$ and therefore leads to $N_{\rm eff} < N_{\rm eff}^{\rm SM}$ \cite{Ibe:2019gpv,Li:2020roy}.

In practice, the constraint from $N_{\rm eff}$ is derived by solving the Boltzmann equations for the coupled SM-dark photon system, taking into account different values of the mixing parameter $\epsilon$ and the dark photon mass $m_{\gamma'}$, as well as the initial dark photon abundance assuming thermal equilibrium at high temperature, as motivated by the ADM scenario. The results are shown in Fig.~\ref{fig:epsilon_vs_mg}, and can be understood from two perspectives.

First, the dark photon must decay before neutrino decoupling, requiring $\tau_{\gamma'} < \mathcal{O}(0.1)\,\mathrm{s}$, thereby setting a lower bound on the parameter space. Second, even if this condition is satisfied, a sufficiently light dark photon may still decay after neutrino decoupling. This latter effect is particularly important, as it imposes a direct lower bound on the dark photon mass $m_{\gamma'}$, largely independent of the value of $\epsilon$, provided that $\epsilon$ is sufficiently large for the dark photon to thermalize with the SM.

The strongest current constraint comes from a combined analysis of the CMB power spectrum, including Planck, ACT, lensing, BAO, and helium/deuterium abundances, yielding $N_{\rm eff} = 2.89 \pm 0.11$ \cite{ACT:2025tim}, compared to the SM prediction $N_{\rm eff}^{\rm SM} = 3.0440 \pm 0.0002$ \cite{Akita:2020szl,Froustey:2020mcq,Bennett:2020zkv,Drewes:2024wbw}. At 95\% C.L., this corresponds to $N_{\rm eff} > 2.67$, which translates into a lower bound on the dark photon mass, $m_{\gamma'}>8.5$ MeV for $\epsilon\gtrsim 10^{-8}$ \cite{Ibe:2019gpv}.

Although the current $N_{\rm eff}$ constraint already excludes roughly half of the parameter space, it could have been significantly stronger if the observed value were closer to the SM prediction. Interestingly, the central value $N_{\rm eff}=2.89$ favors a dark photon mass around $m_{\gamma'} \simeq 12.5~\mathrm{MeV}$. We refer to this alignment as a potential third coincidence, as it falls within the mass range independently suggested by the previous two coincidences. Of course, the present deviation remains statistically consistent with the SM. We anticipate that upcoming CMB upgrades will achieve improved sensitivity, allowing a more precise determination of whether this deviation truly exists. If confirmed, such observations could provide one of the most precise probes of the scenario and substantially narrow down the dark photon mass range.

\section{Direct searches of dark sector}

\subsection{Direct detection of dark matter}

The requirement of a nonzero kinetic mixing parameter $\epsilon$ implies scattering between baryons and dark baryons. Unlike the WIMP with short-range contact interactions, our dark sector involves a light mediator, which necessitates a different treatment \cite{Fornengo:2011sz,Kaplinghat:2013yxa,DelNobile:2015uua}.

With the dark photon mass comparable to the typical momentum transfer, the nuclear recoil interaction becomes momentum dependent, and the signal spectrum peaks toward low recoil energies. Dedicated analyses of low-threshold data from PandaX and XENON \cite{PandaX-II:2018xpz,XENON:2019gfn,PandaX:2023xgl,XENON:2024hup} constrain dark matter with a light mediator scenario in terms of an upper bound on the product $\sigma m_{\gamma'}^4$, where $\sigma$ is the zero-momentum DM–nucleon cross section.

Another distinctive feature of our chiral model is that the dark photon couples only to the axial-vector current $\bar{\psi}\gamma^\mu\gamma^5\psi$, similar to the Majorana dark matter \cite{Chu:2016pew,Chao:2019lhb}. This structure leads to a velocity-suppressed scattering cross section with nuclei, relaxing direct detection constraints significantly compared to conventional SIDM models with vector interactions. In our simplified model, the product is given by
\begin{align}
\sigma m_{\gamma'}^4=
\frac{1}{\pi}\,e^2\,\epsilon^2
\left(\frac{N}{2}\frac{m_{\gamma'}}{f}\right)^2
\left(\frac{Z}{A}\right)^2
\mu_n^2v_0^2~,
\end{align}
where Z and A are the proton number and the mass number
of the nucleus, $\mu_n$ is the DM-nucleon reduced mass, and $v_0$ is the velocity of the dark matter. Such relation allow us to combine the dark photon constraints and the direct detection bound.

Considering the maximal ratio $m_{\gamma'}/m_D = 2.6\times 10^{-3}$, which corresponds to $(\sigma_D/m_D)_{\rm low}= 1.1~\mathrm{cm^2/g}$ and $(\sigma_D/m_D)_{\rm high}= 0.13~\mathrm{cm^2/g}$, the bound $m_{\gamma'}>8.5$ MeV from the $N_{\rm eff}$ constraint translates into a lower bound on the dark matter mass $m_D>3.3$ GeV.

\begin{figure}[tbp]
\centering
\includegraphics[width=0.45\textwidth]{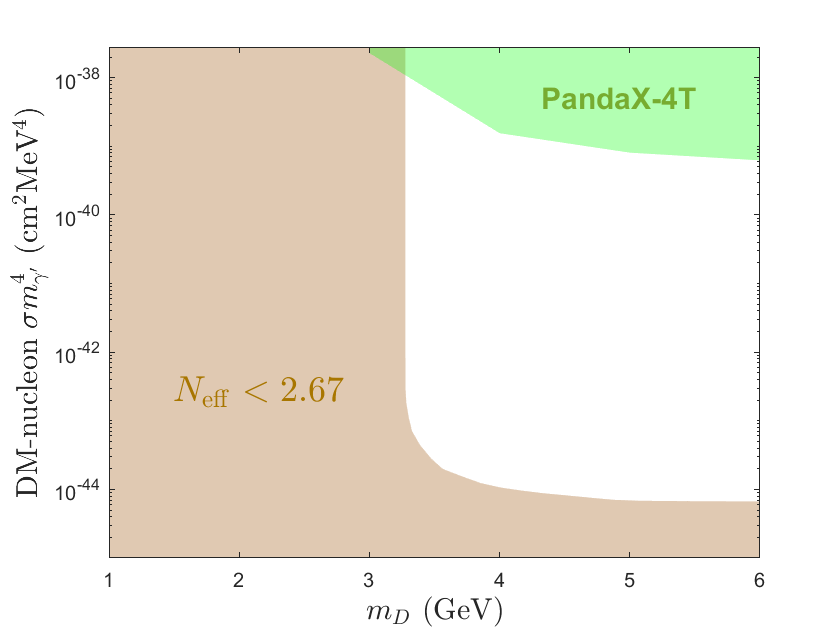}
\caption{Constraints on the $(m_D,\sigma m_{\gamma'}^4)$ plane for the chiral dark QCD model. The green shaded region is excluded by the strongest direct detection bound from PandaX-4T \cite{PandaX:2023xgl}. The brown shaded region is excluded by the $N_{\rm eff}$ bound on dark photon considering the relation in Eq. \eqref{sigmam4}.
\label{fig:sigma_vs_mD}}
\end{figure}

Moreover, the constraint on the dark photon lifetime can be recast into a direct detection bound, as the product can be expressed as
\begin{align}\label{sigmam4}
\sigma m_{\gamma'}^4&=
3.4\times 10^{-45}~{\rm cm^2 \,MeV^4}
\times m_D^\frac{1}{2}\mu_n^2\times\left(\frac{v_0}{10^{-3}}\right)^2\nonumber\\
&\times
\left(\frac{m_{\gamma'}/m_D}{2.6\times 10^{-3}}\right)\left(\frac{0.1~{\rm s}}{\tau_{\gamma'}}\right)
\left(\frac{(\sigma_D/m_D)_{\rm low}}{1.1~\text{cm$^2/$g}}\right)^\frac{1}{2}
\end{align}
This expression explicitly shows how the DM–nucleon cross section depends on the dark photon lifetime and the DM self-interaction cross section. The $N_{\rm eff}$ constraint on the dark photon can therefore be recast as a lower bound on the cross section product as shown in Fig.~\ref{fig:sigma_vs_mD}.

Combined with the current strongest direct detection bound from PandaX-4T \cite{PandaX:2023xgl}, a finite viable parameter space remains, spanning about five orders of magnitude. This remaining region will be probed by future direct detection experiments with improved sensitivity to low recoil energies \cite{SuperCDMS:2018gro,CRESST:2019jnq,GlobalArgonDarkMatter:2022ppc,DarkSide-20k:2024yfq}.

\subsection{Searches of dark photon}

\begin{figure}[tbp]
\centering
\includegraphics[width=0.45\textwidth]{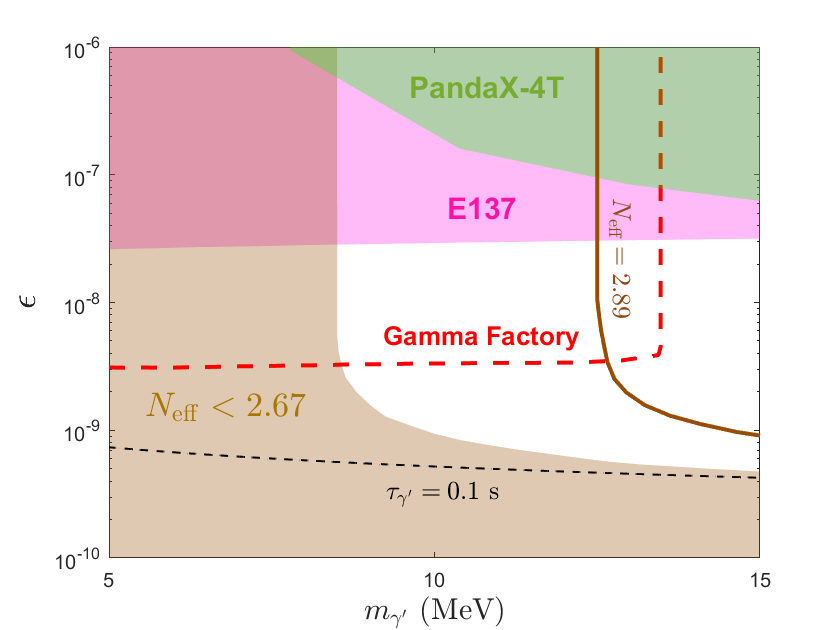}
\caption{Constraints on the dark photon of the chiral dark QCD model with kinetic mixing. The brown shaded region is excluded by the $N_{\rm eff}<2.67$ with a black dashed line corresponding to the lifetime $\tau_{\gamma'}=0.1$ s. The brown line represents the favored values with $N_{\rm eff}=2.89$. The pink shaded region is excluded by the E137 experiment \cite{Bjorken:1988as}. The red dashed line is the projected bound from the Gamma Factory with the energy of $200$ MeV \cite{Chakraborti:2021hfm}. The green shaded region is excluded by the direct detection following the relation in Eq. \eqref{sigmam4}. \label{fig:epsilon_vs_mg}}
\end{figure}

Next, we turn to constraints on the dark photon, focusing on the mass range around $m_{\gamma'} \sim 10$ MeV. In addition to the bound from $N_{\rm eff}$ discussed previously, the relation in Eq.~\eqref{sigmam4} allows us to translate direct detection limits into constraints on the kinetic mixing parameter $\epsilon$, as shown by the green shaded region in Fig. \ref{fig:epsilon_vs_mg}, which already excludes $\epsilon \gtrsim 10^{-7}$.

Searches for dark photons in beam-dump experiments, however, impose even stronger bounds. The current constraint comes from the E137 experiment \cite{Bjorken:1988as}, which restricts $\epsilon < 3\times 10^{-8}$ in the relevant mass range \cite{Bjorken:2009mm,Andreas:2012mt}, as shown by the pink shaded region in Fig.~\ref{fig:epsilon_vs_mg}.

In Fig.~\ref{fig:epsilon_vs_mg}, we do not include the supernova constraints from SN1987A \cite{Chang:2016ntp,Sung:2019xie,DeRocco:2019njg}, as their robustness remains under debate \cite{Bar:2019ifz}. Moreover, the presence of self-interacting dark sector particles may lead to their own self-trapping, potentially rendering these constraints inapplicable \cite{Sung:2021swd}. It has also been shown in Ref.~\cite{Gupta:2025ygk} that the capture of asymmetric dark matter inside the progenitor star can be efficient, suppressing the dark photon luminosity. While supernova constraints on a dark photon–only model may be robust, their applicability to a complete dark sector—particularly one featuring asymmetric or self-interacting dark matter—requires further studies.

There will be more beam dump experiments in future, providing opportunities to probe the relevant parameter space. Among them, the Gamma Factory \cite{Krasny:2015ffb,Krasny:2019wch} is particularly promising. With an expected yield of $3 \times 10^{24}$ photons at 200 MeV, it is projected to probe most of the currently viable parameter space \cite{Chakraborti:2021hfm}, including the particularly interesting region around $m_{\gamma'} \simeq 12.5~\mathrm{MeV}$ suggested by the current $N_{\rm eff}$, with the potential for even broader coverage through further optimization.

\section{Conclusion}

Motivated by two observed coincidences in dark matter measurements, we construct a simplified model of a chiral dark QCD sector, in which the dark baryon serves as dark matter and the dark photon acts both as a light mediator and as a portal to the SM. We analyze the parameter space that can simultaneously accommodate the energy density coincidence and account for the self-interaction cross section required by the core–cusp problem. We find that the preferred region features a dark baryon with mass $m_D=1-5$ GeV and a dark photon with mass $m_{\gamma'}=1-15$ MeV.

Complementary probes are discussed in detail. The most stringent upper bound on the kinetic mixing parameter $\epsilon$ arises from beam-dump searches for dark photons. In contrast, direct detection of dark baryons is less sensitive due to the velocity suppression of the DM–nucleon cross section, a direct consequence of the chiral structure of our dark QCD sector.

The parameter space is bounded from below by measurements of $N_{\rm eff}$, leaving a finite and testable window. Given the strong sensitivity of $N_{\rm eff}$ to MeV-scale new physics, upcoming CMB measurements may provide the first hint of this scenario if the observed value continues to lie below the SM prediction. In parallel, the remaining parameter space can be further explored by the Gamma Factory through dark photon searches, as well as by future direct detection experiments optimized for GeV-scale SIDM with low recoil thresholds.

\section*{Acknowledgments}

I thank Florian Goertz, Ayuki Kamada, and Xiaoyong Chu for useful discussions.

\appendix

\section{An anomaly-free UV Theory}\label{sec:UV}

In the Appendix, we present an example of an anomaly-free UV theory. In the simplified model, there are two Weyl fermions, $\psi_L$ and $\bar{\psi}_R$, transforming as $(\mathbf{N},1/2)$ and $(\overline{\mathbf{N}},1/2)$ under $SU(N)_D \times U(1)_D$. Since they carry the same $U(1)_D$ charge, this setup clearly leads to a $U(1)_D$ gauge anomaly. The anomaly can be canceled by introducing two additional Weyl fermions, $\psi'_L$ and $\bar{\psi}'_R$, transforming as $(\mathbf{N},-1/2)$ and $(\overline{\mathbf{N}},-1/2)$. With these four Weyl fermions, the theory becomes a vector-like QCD theory with $N_f=2$, exhibiting the usual chiral symmetry breaking $SU(2)_L \times SU(2)_R \to SU(2)_V$ and three dark pions.

With only these ingredients, one expect the condensate to form in the neutral channel, $\langle \psi'_L \bar{\psi}_R + \psi'_L \bar{\psi}_R \rangle$, which carries zero net $U(1)_D$ charge, rather than the charged condensate $\Phi \sim \langle \psi_L \bar{\psi}_R \rangle$ required in our simplified model. This is because the gauge interaction favors the unbroken phase. In previous chiral dark sector models \cite{Harigaya:2016rwr,Co:2016akw,Ibe:2021gil}, a fractional charge $a$ was assigned (see Table~\ref{tab:charge1}) so that the condensate always carries a nonzero charge. However, such constructions typically predict a vector coupling between the dark photon and the dark baryon, which is strongly constrained experimentally.

\begin{table}[b]
\centering

\begin{minipage}{0.5\linewidth}
\centering
\begin{tabular}{|c|c|c|}
\hline
                    & $SU(N)_D$   & $U(1)_D$    \\ \hline
$\psi_L $           & $\mathbf{N}$            & $1/2$   \\ 
$\bar{\psi}_{R}$    & $\overline{\mathbf{N}}$ & $-a/2$   \\ 
$\psi'_L $          & $\mathbf{N}$            & $-1/2$   \\ 
$\bar{\psi}'_{R}$   & $\overline{\mathbf{N}}$ & $a/2$   \\ \hline
\end{tabular}
\caption{Particle content of the chiral dark QCD sector in the previous study \cite{Harigaya:2016rwr,Co:2016akw,Ibe:2021gil}, where $0<a<1$ is considered. \label{tab:charge1}}
\end{minipage}
\hfill
\begin{minipage}{0.45\linewidth}
\centering
\begin{tabular}{|c|c|c|}
\hline
                    & $SU(N)_D$   & $U(1)_D$    \\ \hline
$\psi_L $           & $\mathbf{N}$            & $1/2$  \\ 
$\bar{\psi}_{R}$    & $\overline{\mathbf{N}}$ & $1/2$  \\ 
$\psi'_L $          & $\mathbf{N}$            & $-1/2$ \\ 
$\bar{\psi}'_{R}$   & $\overline{\mathbf{N}}$ & $-1/2$ \\ 
$\chi_L $           & $\mathbf{1}$            & $1/2$ \\ 
$\bar{\chi}_{R}$    & ${\mathbf{1}}$ & $1/2$ \\ 
$\chi'_L $          & $\mathbf{1}$            & $-1/2$ \\ 
$\bar{\chi}'_{R}$   & ${\mathbf{1}}$ & $-1/2$ \\ 
\hline
\end{tabular}
\caption{Particle content considered in this study.\label{tab:charge2}}
\end{minipage}
\end{table}

In this study, we retain the original charge assignment and instead aim to realize the desired charged condensate as the true vacuum. To achieve this, we introduce four additional Weyl fermions, $\chi_L$, $\chi'_L$, $\chi_R$, and $\chi'_R$, which are singlets under $SU(N)_D$ but carry the same $U(1)_D$ charges as their corresponding $\psi$ partners. Their charge assignments are summarized in Table~\ref{tab:charge2}.

We further introduce new interactions among fermions mediated by heavy gauge bosons $E_\mu$ and ${E}'_\mu$ with gauge couplings $g_E$ and $g'_E$:
\begin{align}\label{PSint}
{\cal L}_{\text{E}}=~&\frac{1}{\sqrt{2}} g_{E} {E}_{\mu} (\bar{\psi}_L\gamma^\mu \chi_L+\bar{\psi}_R\gamma^\mu \chi_R)\nonumber\\
+~&\frac{1}{\sqrt{2}} g'_{E} {E}'_{\mu} (\bar{\psi}'_L\gamma^\mu \chi'_L+\bar{\psi}'_R\gamma^\mu \chi'_R)\,~,
\end{align}
which lead to effective four-fermion operators given by
\begin{align}
\mathcal{L}_{\text{E,eff}}\supset\frac{g_{E}^2}{M_{E}^2}
\left({{\chi}_R}\bar{\chi}_{L}\right)
\left({{\psi}_L}\bar{\psi}_{R}\right)+
\frac{{g'_E}^2}{{M'_E}^2}
\left({{\chi}'_R}\bar{\chi}'_{L}\right)
\left({{\psi}'_L}\bar{\psi}'_{R}\right)~,
\end{align}
where $M_E$ and $M'_E$ denote the mediator masses. These operators effectively generate Yukawa couplings between $\chi(\chi')$ and the condensates $\Phi(\Phi')$:
\begin{align}
y_\chi\sim 4\pi\left(\frac{f^2}{f_{E}^2}\right)\quad\text{and}\quad
y'_\chi\sim 4\pi\left(\frac{f^2}{{f'_E}^2}\right),~
\end{align}
with $f_E \equiv M_E/g_E$ and $f'_E \equiv M'_E/g'_E$ characterizing the scales of the four-fermion interactions. These interactions favor the charged condensate $\langle \psi_L \bar{\psi}_R + \psi'_L \bar{\psi}'_R \rangle$, where the additional fermions $\chi{(\chi')}$ can acquire masses.

The true vacuum is therefore determined by the competition between the gauge interaction and the four-fermion–induced Yukawa interactions. The desired vacuum $\langle \psi_L \bar{\psi}_R + \psi'_L \bar{\psi}'_R \rangle$ can be realized if
\begin{align}\label{y>g}
y_\chi^2+{y'_\chi}^2>g_D^2~,
\end{align}
which can be satisfied when $f_E$ and $f'_E$ are sufficiently close to the VEV $f$. Similar mechanisms have been studied in other contexts, such as the generation of the top Yukawa coupling in technicolor or composite Higgs models (see, e.g., \cite{Chung:2023iwj}). It has also been shown that a similar IR structure with $f_E, f'_E\sim f$ can arise from a strongly coupled chiral theory \cite{Chung:2024ezq}.

In addition to the extra fermions, the anomaly-free setup features additional pNGBs from the enhanced coset $SU(2)_L \times SU(2)_R / SU(2)_V$. Among the three broken generators, one is eaten by the dark photon, rendering it massive. The remaining two acquire masses from the explicit breaking induced by the four-fermion interactions,
\begin{align}
m^2_{\pi'}\sim \frac{1}{16\pi^2}\left(y_\chi^2+{y'_\chi}^2\right)\Lambda^2
\end{align}
where $\Lambda=4\pi f$. Following the requirement in Eq.~\eqref{y>g}, we find that the same condition gives 
\begin{align}
m_{\pi'}\sim \sqrt{y_\chi^2+{y'_\chi}^2}f>g_Df=m_{\gamma'}.
\end{align}
Since the dark pions are heavier than the dark photon, their effects in cosmology and phenomenology are expected to be subdominant, thereby justifying the use of the simplified model in the main text.

These additional ingredients may also affect the dark baryon mass, analogous to how explicit symmetry-breaking effects—such as quark masses and $U(1)_{\rm EM}$ interactions—modify the proton mass in QCD. Without committing to a specific UV completion, we therefore treat the Yukawa coupling $Y_D$, and consequently the dark baryon mass $m_D$, as free parameters in our analysis.

\bibliography{DM_Ref}

\begin{thebibliography}{102}%
\makeatletter
\providecommand \@ifxundefined [1]{%
 \@ifx{#1\undefined}
}%
\providecommand \@ifnum [1]{%
 \ifnum #1\expandafter \@firstoftwo
 \else \expandafter \@secondoftwo
 \fi
}%
\providecommand \@ifx [1]{%
 \ifx #1\expandafter \@firstoftwo
 \else \expandafter \@secondoftwo
 \fi
}%
\providecommand \natexlab [1]{#1}%
\providecommand \enquote  [1]{``#1''}%
\providecommand \bibnamefont  [1]{#1}%
\providecommand \bibfnamefont [1]{#1}%
\providecommand \citenamefont [1]{#1}%
\providecommand \href@noop [0]{\@secondoftwo}%
\providecommand \href [0]{\begingroup \@sanitize@url \@href}%
\providecommand \@href[1]{\@@startlink{#1}\@@href}%
\providecommand \@@href[1]{\endgroup#1\@@endlink}%
\providecommand \@sanitize@url [0]{\catcode `\\12\catcode `\$12\catcode `\&12\catcode `\#12\catcode `\^12\catcode `\_12\catcode `\%12\relax}%
\providecommand \@@startlink[1]{}%
\providecommand \@@endlink[0]{}%
\providecommand \url  [0]{\begingroup\@sanitize@url \@url }%
\providecommand \@url [1]{\endgroup\@href {#1}{\urlprefix }}%
\providecommand \urlprefix  [0]{URL }%
\providecommand \Eprint [0]{\href }%
\providecommand \doibase [0]{http://dx.doi.org/}%
\providecommand \selectlanguage [0]{\@gobble}%
\providecommand \bibinfo  [0]{\@secondoftwo}%
\providecommand \bibfield  [0]{\@secondoftwo}%
\providecommand \translation [1]{[#1]}%
\providecommand \BibitemOpen [0]{}%
\providecommand \bibitemStop [0]{}%
\providecommand \bibitemNoStop [0]{.\EOS\space}%
\providecommand \EOS [0]{\spacefactor3000\relax}%
\providecommand \BibitemShut  [1]{\csname bibitem#1\endcsname}%
\let\auto@bib@innerbib\@empty
\bibitem [{\citenamefont {Aghanim}\ \emph {et~al.}(2020)\citenamefont {Aghanim} \emph {et~al.}}]{Planck:2018vyg}%
  \BibitemOpen
  \bibfield  {author} {\bibinfo {author} {\bibfnamefont {N.}~\bibnamefont {Aghanim}} \emph {et~al.} (\bibinfo {collaboration} {Planck}),\ }\href {\doibase 10.1051/0004-6361/201833910} {\bibfield  {journal} {\bibinfo  {journal} {Astron. Astrophys.}\ }\textbf {\bibinfo {volume} {641}},\ \bibinfo {pages} {A6} (\bibinfo {year} {2020})},\ \bibinfo {note} {[Erratum: Astron.Astrophys. 652, C4 (2021)]},\ \Eprint {http://arxiv.org/abs/1807.06209} {arXiv:1807.06209 [astro-ph.CO]} \BibitemShut {NoStop}%
\bibitem [{\citenamefont {Hodges}(1993)}]{Hodges:1993yb}%
  \BibitemOpen
  \bibfield  {author} {\bibinfo {author} {\bibfnamefont {H.~M.}\ \bibnamefont {Hodges}},\ }\href {\doibase 10.1103/PhysRevD.47.456} {\bibfield  {journal} {\bibinfo  {journal} {Phys. Rev. D}\ }\textbf {\bibinfo {volume} {47}},\ \bibinfo {pages} {456} (\bibinfo {year} {1993})}\BibitemShut {NoStop}%
\bibitem [{\citenamefont {Berezhiani}\ \emph {et~al.}(1996)\citenamefont {Berezhiani}, \citenamefont {Dolgov},\ and\ \citenamefont {Mohapatra}}]{Berezhiani:1995am}%
  \BibitemOpen
  \bibfield  {author} {\bibinfo {author} {\bibfnamefont {Z.~G.}\ \bibnamefont {Berezhiani}}, \bibinfo {author} {\bibfnamefont {A.~D.}\ \bibnamefont {Dolgov}}, \ and\ \bibinfo {author} {\bibfnamefont {R.~N.}\ \bibnamefont {Mohapatra}},\ }\href {\doibase 10.1016/0370-2693(96)00219-5} {\bibfield  {journal} {\bibinfo  {journal} {Phys. Lett. B}\ }\textbf {\bibinfo {volume} {375}},\ \bibinfo {pages} {26} (\bibinfo {year} {1996})},\ \Eprint {http://arxiv.org/abs/hep-ph/9511221} {arXiv:hep-ph/9511221} \BibitemShut {NoStop}%
\bibitem [{\citenamefont {Foot}\ and\ \citenamefont {Volkas}(2003)}]{Foot:2003jt}%
  \BibitemOpen
  \bibfield  {author} {\bibinfo {author} {\bibfnamefont {R.}~\bibnamefont {Foot}}\ and\ \bibinfo {author} {\bibfnamefont {R.~R.}\ \bibnamefont {Volkas}},\ }\href {\doibase 10.1103/PhysRevD.68.021304} {\bibfield  {journal} {\bibinfo  {journal} {Phys. Rev. D}\ }\textbf {\bibinfo {volume} {68}},\ \bibinfo {pages} {021304} (\bibinfo {year} {2003})},\ \Eprint {http://arxiv.org/abs/hep-ph/0304261} {arXiv:hep-ph/0304261} \BibitemShut {NoStop}%
\bibitem [{\citenamefont {An}\ \emph {et~al.}(2010)\citenamefont {An}, \citenamefont {Chen}, \citenamefont {Mohapatra},\ and\ \citenamefont {Zhang}}]{An:2009vq}%
  \BibitemOpen
  \bibfield  {author} {\bibinfo {author} {\bibfnamefont {H.}~\bibnamefont {An}}, \bibinfo {author} {\bibfnamefont {S.-L.}\ \bibnamefont {Chen}}, \bibinfo {author} {\bibfnamefont {R.~N.}\ \bibnamefont {Mohapatra}}, \ and\ \bibinfo {author} {\bibfnamefont {Y.}~\bibnamefont {Zhang}},\ }\href {\doibase 10.1007/JHEP03(2010)124} {\bibfield  {journal} {\bibinfo  {journal} {JHEP}\ }\textbf {\bibinfo {volume} {03}},\ \bibinfo {pages} {124} (\bibinfo {year} {2010})},\ \Eprint {http://arxiv.org/abs/0911.4463} {arXiv:0911.4463 [hep-ph]} \BibitemShut {NoStop}%
\bibitem [{\citenamefont {Farina}(2015)}]{Farina:2015uea}%
  \BibitemOpen
  \bibfield  {author} {\bibinfo {author} {\bibfnamefont {M.}~\bibnamefont {Farina}},\ }\href {\doibase 10.1088/1475-7516/2015/11/017} {\bibfield  {journal} {\bibinfo  {journal} {JCAP}\ }\textbf {\bibinfo {volume} {11}},\ \bibinfo {pages} {017} (\bibinfo {year} {2015})},\ \Eprint {http://arxiv.org/abs/1506.03520} {arXiv:1506.03520 [hep-ph]} \BibitemShut {NoStop}%
\bibitem [{\citenamefont {Garcia~Garcia}\ \emph {et~al.}(2015)\citenamefont {Garcia~Garcia}, \citenamefont {Lasenby},\ and\ \citenamefont {March-Russell}}]{GarciaGarcia:2015pnn}%
  \BibitemOpen
  \bibfield  {author} {\bibinfo {author} {\bibfnamefont {I.}~\bibnamefont {Garcia~Garcia}}, \bibinfo {author} {\bibfnamefont {R.}~\bibnamefont {Lasenby}}, \ and\ \bibinfo {author} {\bibfnamefont {J.}~\bibnamefont {March-Russell}},\ }\href {\doibase 10.1103/PhysRevLett.115.121801} {\bibfield  {journal} {\bibinfo  {journal} {Phys. Rev. Lett.}\ }\textbf {\bibinfo {volume} {115}},\ \bibinfo {pages} {121801} (\bibinfo {year} {2015})},\ \Eprint {http://arxiv.org/abs/1505.07410} {arXiv:1505.07410 [hep-ph]} \BibitemShut {NoStop}%
\bibitem [{\citenamefont {Lonsdale}\ and\ \citenamefont {Volkas}(2018)}]{Lonsdale:2018xwd}%
  \BibitemOpen
  \bibfield  {author} {\bibinfo {author} {\bibfnamefont {S.~J.}\ \bibnamefont {Lonsdale}}\ and\ \bibinfo {author} {\bibfnamefont {R.~R.}\ \bibnamefont {Volkas}},\ }\href {\doibase 10.1103/PhysRevD.97.103510} {\bibfield  {journal} {\bibinfo  {journal} {Phys. Rev. D}\ }\textbf {\bibinfo {volume} {97}},\ \bibinfo {pages} {103510} (\bibinfo {year} {2018})},\ \Eprint {http://arxiv.org/abs/1801.05561} {arXiv:1801.05561 [hep-ph]} \BibitemShut {NoStop}%
\bibitem [{\citenamefont {Ibe}\ \emph {et~al.}(2019)\citenamefont {Ibe}, \citenamefont {Kamada}, \citenamefont {Kobayashi}, \citenamefont {Kuwahara},\ and\ \citenamefont {Nakano}}]{Ibe:2019ena}%
  \BibitemOpen
  \bibfield  {author} {\bibinfo {author} {\bibfnamefont {M.}~\bibnamefont {Ibe}}, \bibinfo {author} {\bibfnamefont {A.}~\bibnamefont {Kamada}}, \bibinfo {author} {\bibfnamefont {S.}~\bibnamefont {Kobayashi}}, \bibinfo {author} {\bibfnamefont {T.}~\bibnamefont {Kuwahara}}, \ and\ \bibinfo {author} {\bibfnamefont {W.}~\bibnamefont {Nakano}},\ }\href {\doibase 10.1103/PhysRevD.100.075022} {\bibfield  {journal} {\bibinfo  {journal} {Phys. Rev. D}\ }\textbf {\bibinfo {volume} {100}},\ \bibinfo {pages} {075022} (\bibinfo {year} {2019})},\ \Eprint {http://arxiv.org/abs/1907.03404} {arXiv:1907.03404 [hep-ph]} \BibitemShut {NoStop}%
\bibitem [{\citenamefont {Bodas}\ \emph {et~al.}(2024)\citenamefont {Bodas}, \citenamefont {Buen-Abad}, \citenamefont {Hook},\ and\ \citenamefont {Sundrum}}]{Bodas:2024idn}%
  \BibitemOpen
  \bibfield  {author} {\bibinfo {author} {\bibfnamefont {A.}~\bibnamefont {Bodas}}, \bibinfo {author} {\bibfnamefont {M.~A.}\ \bibnamefont {Buen-Abad}}, \bibinfo {author} {\bibfnamefont {A.}~\bibnamefont {Hook}}, \ and\ \bibinfo {author} {\bibfnamefont {R.}~\bibnamefont {Sundrum}},\ }\href@noop {} {\  (\bibinfo {year} {2024})},\ \Eprint {http://arxiv.org/abs/2401.12286} {arXiv:2401.12286 [hep-ph]} \BibitemShut {NoStop}%
\bibitem [{\citenamefont {Murgui}\ and\ \citenamefont {Zurek}(2022)}]{Murgui:2021eqf}%
  \BibitemOpen
  \bibfield  {author} {\bibinfo {author} {\bibfnamefont {C.}~\bibnamefont {Murgui}}\ and\ \bibinfo {author} {\bibfnamefont {K.~M.}\ \bibnamefont {Zurek}},\ }\href {\doibase 10.1103/PhysRevD.105.095002} {\bibfield  {journal} {\bibinfo  {journal} {Phys. Rev. D}\ }\textbf {\bibinfo {volume} {105}},\ \bibinfo {pages} {095002} (\bibinfo {year} {2022})},\ \Eprint {http://arxiv.org/abs/2112.08374} {arXiv:2112.08374 [hep-ph]} \BibitemShut {NoStop}%
\bibitem [{\citenamefont {Chung}(2024{\natexlab{a}})}]{Chung:2024nnj}%
  \BibitemOpen
  \bibfield  {author} {\bibinfo {author} {\bibfnamefont {Y.}~\bibnamefont {Chung}},\ }\href@noop {} {\  (\bibinfo {year} {2024}{\natexlab{a}})},\ \Eprint {http://arxiv.org/abs/2411.16860} {arXiv:2411.16860 [hep-ph]} \BibitemShut {NoStop}%
\bibitem [{\citenamefont {Bai}\ and\ \citenamefont {Schwaller}(2014)}]{Bai:2013xga}%
  \BibitemOpen
  \bibfield  {author} {\bibinfo {author} {\bibfnamefont {Y.}~\bibnamefont {Bai}}\ and\ \bibinfo {author} {\bibfnamefont {P.}~\bibnamefont {Schwaller}},\ }\href {\doibase 10.1103/PhysRevD.89.063522} {\bibfield  {journal} {\bibinfo  {journal} {Phys. Rev. D}\ }\textbf {\bibinfo {volume} {89}},\ \bibinfo {pages} {063522} (\bibinfo {year} {2014})},\ \Eprint {http://arxiv.org/abs/1306.4676} {arXiv:1306.4676 [hep-ph]} \BibitemShut {NoStop}%
\bibitem [{\citenamefont {Newstead}\ and\ \citenamefont {TerBeek}(2014)}]{Newstead:2014jva}%
  \BibitemOpen
  \bibfield  {author} {\bibinfo {author} {\bibfnamefont {J.~L.}\ \bibnamefont {Newstead}}\ and\ \bibinfo {author} {\bibfnamefont {R.~H.}\ \bibnamefont {TerBeek}},\ }\href {\doibase 10.1103/PhysRevD.90.074008} {\bibfield  {journal} {\bibinfo  {journal} {Phys. Rev. D}\ }\textbf {\bibinfo {volume} {90}},\ \bibinfo {pages} {074008} (\bibinfo {year} {2014})},\ \Eprint {http://arxiv.org/abs/1405.7427} {arXiv:1405.7427 [hep-ph]} \BibitemShut {NoStop}%
\bibitem [{\citenamefont {Ritter}\ and\ \citenamefont {Volkas}(2023)}]{Ritter:2022opo}%
  \BibitemOpen
  \bibfield  {author} {\bibinfo {author} {\bibfnamefont {A.~C.}\ \bibnamefont {Ritter}}\ and\ \bibinfo {author} {\bibfnamefont {R.~R.}\ \bibnamefont {Volkas}},\ }\href {\doibase 10.1103/PhysRevD.107.015029} {\bibfield  {journal} {\bibinfo  {journal} {Phys. Rev. D}\ }\textbf {\bibinfo {volume} {107}},\ \bibinfo {pages} {015029} (\bibinfo {year} {2023})},\ \Eprint {http://arxiv.org/abs/2210.11011} {arXiv:2210.11011 [hep-ph]} \BibitemShut {NoStop}%
\bibitem [{\citenamefont {Ritter}\ and\ \citenamefont {Volkas}(2024)}]{Ritter:2024sqv}%
  \BibitemOpen
  \bibfield  {author} {\bibinfo {author} {\bibfnamefont {A.~C.}\ \bibnamefont {Ritter}}\ and\ \bibinfo {author} {\bibfnamefont {R.~R.}\ \bibnamefont {Volkas}},\ }\href@noop {} {\  (\bibinfo {year} {2024})},\ \Eprint {http://arxiv.org/abs/2404.05999} {arXiv:2404.05999 [hep-ph]} \BibitemShut {NoStop}%
\bibitem [{\citenamefont {Chung}(2024{\natexlab{b}})}]{Chung:2024ezq}%
  \BibitemOpen
  \bibfield  {author} {\bibinfo {author} {\bibfnamefont {Y.}~\bibnamefont {Chung}},\ }\href@noop {} {\  (\bibinfo {year} {2024}{\natexlab{b}})},\ \Eprint {http://arxiv.org/abs/2411.18725} {arXiv:2411.18725 [hep-ph]} \BibitemShut {NoStop}%
\bibitem [{\citenamefont {Spergel}\ and\ \citenamefont {Press}(1985)}]{Spergel:1984re}%
  \BibitemOpen
  \bibfield  {author} {\bibinfo {author} {\bibfnamefont {D.~N.}\ \bibnamefont {Spergel}}\ and\ \bibinfo {author} {\bibfnamefont {W.~H.}\ \bibnamefont {Press}},\ }\href {\doibase 10.1086/163336} {\bibfield  {journal} {\bibinfo  {journal} {Astrophys. J.}\ }\textbf {\bibinfo {volume} {294}},\ \bibinfo {pages} {663} (\bibinfo {year} {1985})}\BibitemShut {NoStop}%
\bibitem [{\citenamefont {Nussinov}(1985)}]{Nussinov:1985xr}%
  \BibitemOpen
  \bibfield  {author} {\bibinfo {author} {\bibfnamefont {S.}~\bibnamefont {Nussinov}},\ }\href {\doibase 10.1016/0370-2693(85)90689-6} {\bibfield  {journal} {\bibinfo  {journal} {Phys. Lett. B}\ }\textbf {\bibinfo {volume} {165}},\ \bibinfo {pages} {55} (\bibinfo {year} {1985})}\BibitemShut {NoStop}%
\bibitem [{\citenamefont {Gelmini}\ \emph {et~al.}(1987)\citenamefont {Gelmini}, \citenamefont {Hall},\ and\ \citenamefont {Lin}}]{Gelmini:1986zz}%
  \BibitemOpen
  \bibfield  {author} {\bibinfo {author} {\bibfnamefont {G.~B.}\ \bibnamefont {Gelmini}}, \bibinfo {author} {\bibfnamefont {L.~J.}\ \bibnamefont {Hall}}, \ and\ \bibinfo {author} {\bibfnamefont {M.~J.}\ \bibnamefont {Lin}},\ }\href {\doibase 10.1016/0550-3213(87)90424-X} {\bibfield  {journal} {\bibinfo  {journal} {Nucl. Phys. B}\ }\textbf {\bibinfo {volume} {281}},\ \bibinfo {pages} {726} (\bibinfo {year} {1987})}\BibitemShut {NoStop}%
\bibitem [{\citenamefont {Barr}\ \emph {et~al.}(1990)\citenamefont {Barr}, \citenamefont {Chivukula},\ and\ \citenamefont {Farhi}}]{Barr:1990ca}%
  \BibitemOpen
  \bibfield  {author} {\bibinfo {author} {\bibfnamefont {S.~M.}\ \bibnamefont {Barr}}, \bibinfo {author} {\bibfnamefont {R.~S.}\ \bibnamefont {Chivukula}}, \ and\ \bibinfo {author} {\bibfnamefont {E.}~\bibnamefont {Farhi}},\ }\href {\doibase 10.1016/0370-2693(90)91661-T} {\bibfield  {journal} {\bibinfo  {journal} {Phys. Lett. B}\ }\textbf {\bibinfo {volume} {241}},\ \bibinfo {pages} {387} (\bibinfo {year} {1990})}\BibitemShut {NoStop}%
\bibitem [{\citenamefont {Barr}(1991)}]{Barr:1991qn}%
  \BibitemOpen
  \bibfield  {author} {\bibinfo {author} {\bibfnamefont {S.~M.}\ \bibnamefont {Barr}},\ }\href {\doibase 10.1103/PhysRevD.44.3062} {\bibfield  {journal} {\bibinfo  {journal} {Phys. Rev. D}\ }\textbf {\bibinfo {volume} {44}},\ \bibinfo {pages} {3062} (\bibinfo {year} {1991})}\BibitemShut {NoStop}%
\bibitem [{\citenamefont {Kaplan}(1992)}]{Kaplan:1991ah}%
  \BibitemOpen
  \bibfield  {author} {\bibinfo {author} {\bibfnamefont {D.~B.}\ \bibnamefont {Kaplan}},\ }\href {\doibase 10.1103/PhysRevLett.68.741} {\bibfield  {journal} {\bibinfo  {journal} {Phys. Rev. Lett.}\ }\textbf {\bibinfo {volume} {68}},\ \bibinfo {pages} {741} (\bibinfo {year} {1992})}\BibitemShut {NoStop}%
\bibitem [{\citenamefont {Gudnason}\ \emph {et~al.}(2006{\natexlab{a}})\citenamefont {Gudnason}, \citenamefont {Kouvaris},\ and\ \citenamefont {Sannino}}]{Gudnason:2006ug}%
  \BibitemOpen
  \bibfield  {author} {\bibinfo {author} {\bibfnamefont {S.~B.}\ \bibnamefont {Gudnason}}, \bibinfo {author} {\bibfnamefont {C.}~\bibnamefont {Kouvaris}}, \ and\ \bibinfo {author} {\bibfnamefont {F.}~\bibnamefont {Sannino}},\ }\href {\doibase 10.1103/PhysRevD.73.115003} {\bibfield  {journal} {\bibinfo  {journal} {Phys. Rev. D}\ }\textbf {\bibinfo {volume} {73}},\ \bibinfo {pages} {115003} (\bibinfo {year} {2006}{\natexlab{a}})},\ \Eprint {http://arxiv.org/abs/hep-ph/0603014} {arXiv:hep-ph/0603014} \BibitemShut {NoStop}%
\bibitem [{\citenamefont {Gudnason}\ \emph {et~al.}(2006{\natexlab{b}})\citenamefont {Gudnason}, \citenamefont {Kouvaris},\ and\ \citenamefont {Sannino}}]{Gudnason:2006yj}%
  \BibitemOpen
  \bibfield  {author} {\bibinfo {author} {\bibfnamefont {S.~B.}\ \bibnamefont {Gudnason}}, \bibinfo {author} {\bibfnamefont {C.}~\bibnamefont {Kouvaris}}, \ and\ \bibinfo {author} {\bibfnamefont {F.}~\bibnamefont {Sannino}},\ }\href {\doibase 10.1103/PhysRevD.74.095008} {\bibfield  {journal} {\bibinfo  {journal} {Phys. Rev. D}\ }\textbf {\bibinfo {volume} {74}},\ \bibinfo {pages} {095008} (\bibinfo {year} {2006}{\natexlab{b}})},\ \Eprint {http://arxiv.org/abs/hep-ph/0608055} {arXiv:hep-ph/0608055} \BibitemShut {NoStop}%
\bibitem [{\citenamefont {Kaplan}\ \emph {et~al.}(2009)\citenamefont {Kaplan}, \citenamefont {Luty},\ and\ \citenamefont {Zurek}}]{Kaplan:2009ag}%
  \BibitemOpen
  \bibfield  {author} {\bibinfo {author} {\bibfnamefont {D.~E.}\ \bibnamefont {Kaplan}}, \bibinfo {author} {\bibfnamefont {M.~A.}\ \bibnamefont {Luty}}, \ and\ \bibinfo {author} {\bibfnamefont {K.~M.}\ \bibnamefont {Zurek}},\ }\href {\doibase 10.1103/PhysRevD.79.115016} {\bibfield  {journal} {\bibinfo  {journal} {Phys. Rev. D}\ }\textbf {\bibinfo {volume} {79}},\ \bibinfo {pages} {115016} (\bibinfo {year} {2009})},\ \Eprint {http://arxiv.org/abs/0901.4117} {arXiv:0901.4117 [hep-ph]} \BibitemShut {NoStop}%
\bibitem [{\citenamefont {Davoudiasl}\ and\ \citenamefont {Mohapatra}(2012)}]{Davoudiasl:2012uw}%
  \BibitemOpen
  \bibfield  {author} {\bibinfo {author} {\bibfnamefont {H.}~\bibnamefont {Davoudiasl}}\ and\ \bibinfo {author} {\bibfnamefont {R.~N.}\ \bibnamefont {Mohapatra}},\ }\href {\doibase 10.1088/1367-2630/14/9/095011} {\bibfield  {journal} {\bibinfo  {journal} {New J. Phys.}\ }\textbf {\bibinfo {volume} {14}},\ \bibinfo {pages} {095011} (\bibinfo {year} {2012})},\ \Eprint {http://arxiv.org/abs/1203.1247} {arXiv:1203.1247 [hep-ph]} \BibitemShut {NoStop}%
\bibitem [{\citenamefont {Petraki}\ and\ \citenamefont {Volkas}(2013)}]{Petraki:2013wwa}%
  \BibitemOpen
  \bibfield  {author} {\bibinfo {author} {\bibfnamefont {K.}~\bibnamefont {Petraki}}\ and\ \bibinfo {author} {\bibfnamefont {R.~R.}\ \bibnamefont {Volkas}},\ }\href {\doibase 10.1142/S0217751X13300287} {\bibfield  {journal} {\bibinfo  {journal} {Int. J. Mod. Phys. A}\ }\textbf {\bibinfo {volume} {28}},\ \bibinfo {pages} {1330028} (\bibinfo {year} {2013})},\ \Eprint {http://arxiv.org/abs/1305.4939} {arXiv:1305.4939 [hep-ph]} \BibitemShut {NoStop}%
\bibitem [{\citenamefont {Zurek}(2014)}]{Zurek:2013wia}%
  \BibitemOpen
  \bibfield  {author} {\bibinfo {author} {\bibfnamefont {K.~M.}\ \bibnamefont {Zurek}},\ }\href {\doibase 10.1016/j.physrep.2013.12.001} {\bibfield  {journal} {\bibinfo  {journal} {Phys. Rept.}\ }\textbf {\bibinfo {volume} {537}},\ \bibinfo {pages} {91} (\bibinfo {year} {2014})},\ \Eprint {http://arxiv.org/abs/1308.0338} {arXiv:1308.0338 [hep-ph]} \BibitemShut {NoStop}%
\bibitem [{\citenamefont {Rosa}\ and\ \citenamefont {Silva}(2023)}]{Rosa:2022sym}%
  \BibitemOpen
  \bibfield  {author} {\bibinfo {author} {\bibfnamefont {J.~a.~G.}\ \bibnamefont {Rosa}}\ and\ \bibinfo {author} {\bibfnamefont {D.~M.~C.}\ \bibnamefont {Silva}},\ }\href {\doibase 10.1016/j.physletb.2023.138178} {\bibfield  {journal} {\bibinfo  {journal} {Phys. Lett. B}\ }\textbf {\bibinfo {volume} {846}},\ \bibinfo {pages} {138178} (\bibinfo {year} {2023})},\ \Eprint {http://arxiv.org/abs/2211.10359} {arXiv:2211.10359 [hep-ph]} \BibitemShut {NoStop}%
\bibitem [{\citenamefont {Brzeminski}\ and\ \citenamefont {Hook}(2024)}]{Brzeminski:2023wza}%
  \BibitemOpen
  \bibfield  {author} {\bibinfo {author} {\bibfnamefont {D.}~\bibnamefont {Brzeminski}}\ and\ \bibinfo {author} {\bibfnamefont {A.}~\bibnamefont {Hook}},\ }\href {\doibase 10.1103/PhysRevLett.132.201001} {\bibfield  {journal} {\bibinfo  {journal} {Phys. Rev. Lett.}\ }\textbf {\bibinfo {volume} {132}},\ \bibinfo {pages} {201001} (\bibinfo {year} {2024})},\ \Eprint {http://arxiv.org/abs/2310.07777} {arXiv:2310.07777 [hep-ph]} \BibitemShut {NoStop}%
\bibitem [{\citenamefont {Banerjee}\ \emph {et~al.}(2024)\citenamefont {Banerjee}, \citenamefont {Brzeminski},\ and\ \citenamefont {Hook}}]{Banerjee:2024xhn}%
  \BibitemOpen
  \bibfield  {author} {\bibinfo {author} {\bibfnamefont {A.}~\bibnamefont {Banerjee}}, \bibinfo {author} {\bibfnamefont {D.}~\bibnamefont {Brzeminski}}, \ and\ \bibinfo {author} {\bibfnamefont {A.}~\bibnamefont {Hook}},\ }\href@noop {} {\  (\bibinfo {year} {2024})},\ \Eprint {http://arxiv.org/abs/2410.22412} {arXiv:2410.22412 [hep-ph]} \BibitemShut {NoStop}%
\bibitem [{\citenamefont {Flores}\ and\ \citenamefont {Primack}(1994)}]{Flores:1994gz}%
  \BibitemOpen
  \bibfield  {author} {\bibinfo {author} {\bibfnamefont {R.~A.}\ \bibnamefont {Flores}}\ and\ \bibinfo {author} {\bibfnamefont {J.~R.}\ \bibnamefont {Primack}},\ }\href {\doibase 10.1086/187350} {\bibfield  {journal} {\bibinfo  {journal} {Astrophys. J. Lett.}\ }\textbf {\bibinfo {volume} {427}},\ \bibinfo {pages} {L1} (\bibinfo {year} {1994})},\ \Eprint {http://arxiv.org/abs/astro-ph/9402004} {arXiv:astro-ph/9402004} \BibitemShut {NoStop}%
\bibitem [{\citenamefont {Moore}(1994)}]{Moore:1994yx}%
  \BibitemOpen
  \bibfield  {author} {\bibinfo {author} {\bibfnamefont {B.}~\bibnamefont {Moore}},\ }\href {\doibase 10.1038/370629a0} {\bibfield  {journal} {\bibinfo  {journal} {Nature}\ }\textbf {\bibinfo {volume} {370}},\ \bibinfo {pages} {629} (\bibinfo {year} {1994})}\BibitemShut {NoStop}%
\bibitem [{\citenamefont {Moore}\ \emph {et~al.}(1999)\citenamefont {Moore}, \citenamefont {Quinn}, \citenamefont {Governato}, \citenamefont {Stadel},\ and\ \citenamefont {Lake}}]{Moore:1999gc}%
  \BibitemOpen
  \bibfield  {author} {\bibinfo {author} {\bibfnamefont {B.}~\bibnamefont {Moore}}, \bibinfo {author} {\bibfnamefont {T.~R.}\ \bibnamefont {Quinn}}, \bibinfo {author} {\bibfnamefont {F.}~\bibnamefont {Governato}}, \bibinfo {author} {\bibfnamefont {J.}~\bibnamefont {Stadel}}, \ and\ \bibinfo {author} {\bibfnamefont {G.}~\bibnamefont {Lake}},\ }\href {\doibase 10.1046/j.1365-8711.1999.03039.x} {\bibfield  {journal} {\bibinfo  {journal} {Mon. Not. Roy. Astron. Soc.}\ }\textbf {\bibinfo {volume} {310}},\ \bibinfo {pages} {1147} (\bibinfo {year} {1999})},\ \Eprint {http://arxiv.org/abs/astro-ph/9903164} {arXiv:astro-ph/9903164} \BibitemShut {NoStop}%
\bibitem [{\citenamefont {Walker}\ and\ \citenamefont {Penarrubia}(2011)}]{Walker:2011zu}%
  \BibitemOpen
  \bibfield  {author} {\bibinfo {author} {\bibfnamefont {M.~G.}\ \bibnamefont {Walker}}\ and\ \bibinfo {author} {\bibfnamefont {J.}~\bibnamefont {Penarrubia}},\ }\href {\doibase 10.1088/0004-637X/742/1/20} {\bibfield  {journal} {\bibinfo  {journal} {Astrophys. J.}\ }\textbf {\bibinfo {volume} {742}},\ \bibinfo {pages} {20} (\bibinfo {year} {2011})},\ \Eprint {http://arxiv.org/abs/1108.2404} {arXiv:1108.2404 [astro-ph.CO]} \BibitemShut {NoStop}%
\bibitem [{\citenamefont {Tollerud}\ \emph {et~al.}(2014)\citenamefont {Tollerud}, \citenamefont {Boylan-Kolchin},\ and\ \citenamefont {Bullock}}]{Tollerud:2014zha}%
  \BibitemOpen
  \bibfield  {author} {\bibinfo {author} {\bibfnamefont {E.~J.}\ \bibnamefont {Tollerud}}, \bibinfo {author} {\bibfnamefont {M.}~\bibnamefont {Boylan-Kolchin}}, \ and\ \bibinfo {author} {\bibfnamefont {J.~S.}\ \bibnamefont {Bullock}},\ }\href {\doibase 10.1093/mnras/stu474} {\bibfield  {journal} {\bibinfo  {journal} {Mon. Not. Roy. Astron. Soc.}\ }\textbf {\bibinfo {volume} {440}},\ \bibinfo {pages} {3511} (\bibinfo {year} {2014})},\ \Eprint {http://arxiv.org/abs/1403.6469} {arXiv:1403.6469 [astro-ph.GA]} \BibitemShut {NoStop}%
\bibitem [{\citenamefont {de~Blok}\ \emph {et~al.}(2001)\citenamefont {de~Blok}, \citenamefont {McGaugh}, \citenamefont {Bosma},\ and\ \citenamefont {Rubin}}]{deBlok:2001hbg}%
  \BibitemOpen
  \bibfield  {author} {\bibinfo {author} {\bibfnamefont {W.~J.~G.}\ \bibnamefont {de~Blok}}, \bibinfo {author} {\bibfnamefont {S.~S.}\ \bibnamefont {McGaugh}}, \bibinfo {author} {\bibfnamefont {A.}~\bibnamefont {Bosma}}, \ and\ \bibinfo {author} {\bibfnamefont {V.~C.}\ \bibnamefont {Rubin}},\ }\href {\doibase 10.1086/320262} {\bibfield  {journal} {\bibinfo  {journal} {Astrophys. J. Lett.}\ }\textbf {\bibinfo {volume} {552}},\ \bibinfo {pages} {L23} (\bibinfo {year} {2001})},\ \Eprint {http://arxiv.org/abs/astro-ph/0103102} {arXiv:astro-ph/0103102} \BibitemShut {NoStop}%
\bibitem [{\citenamefont {de~Blok}\ and\ \citenamefont {Bosma}(2002)}]{deBlok:2002vgq}%
  \BibitemOpen
  \bibfield  {author} {\bibinfo {author} {\bibfnamefont {W.~J.~G.}\ \bibnamefont {de~Blok}}\ and\ \bibinfo {author} {\bibfnamefont {A.}~\bibnamefont {Bosma}},\ }\href {\doibase 10.1051/0004-6361:20020080} {\bibfield  {journal} {\bibinfo  {journal} {Astron. Astrophys.}\ }\textbf {\bibinfo {volume} {385}},\ \bibinfo {pages} {816} (\bibinfo {year} {2002})},\ \Eprint {http://arxiv.org/abs/astro-ph/0201276} {arXiv:astro-ph/0201276} \BibitemShut {NoStop}%
\bibitem [{\citenamefont {Simon}\ \emph {et~al.}(2005)\citenamefont {Simon}, \citenamefont {Bolatto}, \citenamefont {Leroy}, \citenamefont {Blitz},\ and\ \citenamefont {Gates}}]{Simon:2004sr}%
  \BibitemOpen
  \bibfield  {author} {\bibinfo {author} {\bibfnamefont {J.~D.}\ \bibnamefont {Simon}}, \bibinfo {author} {\bibfnamefont {A.~D.}\ \bibnamefont {Bolatto}}, \bibinfo {author} {\bibfnamefont {A.}~\bibnamefont {Leroy}}, \bibinfo {author} {\bibfnamefont {L.}~\bibnamefont {Blitz}}, \ and\ \bibinfo {author} {\bibfnamefont {E.~L.}\ \bibnamefont {Gates}},\ }\href {\doibase 10.1086/427684} {\bibfield  {journal} {\bibinfo  {journal} {Astrophys. J.}\ }\textbf {\bibinfo {volume} {621}},\ \bibinfo {pages} {757} (\bibinfo {year} {2005})},\ \Eprint {http://arxiv.org/abs/astro-ph/0412035} {arXiv:astro-ph/0412035} \BibitemShut {NoStop}%
\bibitem [{\citenamefont {Gentile}\ \emph {et~al.}(2004)\citenamefont {Gentile}, \citenamefont {Salucci}, \citenamefont {Klein}, \citenamefont {Vergani},\ and\ \citenamefont {Kalberla}}]{Gentile:2004tb}%
  \BibitemOpen
  \bibfield  {author} {\bibinfo {author} {\bibfnamefont {G.}~\bibnamefont {Gentile}}, \bibinfo {author} {\bibfnamefont {P.}~\bibnamefont {Salucci}}, \bibinfo {author} {\bibfnamefont {U.}~\bibnamefont {Klein}}, \bibinfo {author} {\bibfnamefont {D.}~\bibnamefont {Vergani}}, \ and\ \bibinfo {author} {\bibfnamefont {P.}~\bibnamefont {Kalberla}},\ }\href {\doibase 10.1111/j.1365-2966.2004.07836.x} {\bibfield  {journal} {\bibinfo  {journal} {Mon. Not. Roy. Astron. Soc.}\ }\textbf {\bibinfo {volume} {351}},\ \bibinfo {pages} {903} (\bibinfo {year} {2004})},\ \Eprint {http://arxiv.org/abs/astro-ph/0403154} {arXiv:astro-ph/0403154} \BibitemShut {NoStop}%
\bibitem [{\citenamefont {Newman}\ \emph {et~al.}(2013{\natexlab{a}})\citenamefont {Newman}, \citenamefont {Treu}, \citenamefont {Ellis}, \citenamefont {Sand}, \citenamefont {Nipoti}, \citenamefont {Richard},\ and\ \citenamefont {Jullo}}]{Newman:2012nv}%
  \BibitemOpen
  \bibfield  {author} {\bibinfo {author} {\bibfnamefont {A.~B.}\ \bibnamefont {Newman}}, \bibinfo {author} {\bibfnamefont {T.}~\bibnamefont {Treu}}, \bibinfo {author} {\bibfnamefont {R.~S.}\ \bibnamefont {Ellis}}, \bibinfo {author} {\bibfnamefont {D.~J.}\ \bibnamefont {Sand}}, \bibinfo {author} {\bibfnamefont {C.}~\bibnamefont {Nipoti}}, \bibinfo {author} {\bibfnamefont {J.}~\bibnamefont {Richard}}, \ and\ \bibinfo {author} {\bibfnamefont {E.}~\bibnamefont {Jullo}},\ }\href {\doibase 10.1088/0004-637X/765/1/24} {\bibfield  {journal} {\bibinfo  {journal} {Astrophys. J.}\ }\textbf {\bibinfo {volume} {765}},\ \bibinfo {pages} {24} (\bibinfo {year} {2013}{\natexlab{a}})},\ \Eprint {http://arxiv.org/abs/1209.1391} {arXiv:1209.1391 [astro-ph.CO]} \BibitemShut {NoStop}%
\bibitem [{\citenamefont {Newman}\ \emph {et~al.}(2013{\natexlab{b}})\citenamefont {Newman}, \citenamefont {Treu}, \citenamefont {Ellis},\ and\ \citenamefont {Sand}}]{Newman:2012nw}%
  \BibitemOpen
  \bibfield  {author} {\bibinfo {author} {\bibfnamefont {A.~B.}\ \bibnamefont {Newman}}, \bibinfo {author} {\bibfnamefont {T.}~\bibnamefont {Treu}}, \bibinfo {author} {\bibfnamefont {R.~S.}\ \bibnamefont {Ellis}}, \ and\ \bibinfo {author} {\bibfnamefont {D.~J.}\ \bibnamefont {Sand}},\ }\href {\doibase 10.1088/0004-637X/765/1/25} {\bibfield  {journal} {\bibinfo  {journal} {Astrophys. J.}\ }\textbf {\bibinfo {volume} {765}},\ \bibinfo {pages} {25} (\bibinfo {year} {2013}{\natexlab{b}})},\ \Eprint {http://arxiv.org/abs/1209.1392} {arXiv:1209.1392 [astro-ph.CO]} \BibitemShut {NoStop}%
\bibitem [{\citenamefont {Dubinski}\ and\ \citenamefont {Carlberg}(1991)}]{Dubinski:1991bm}%
  \BibitemOpen
  \bibfield  {author} {\bibinfo {author} {\bibfnamefont {J.}~\bibnamefont {Dubinski}}\ and\ \bibinfo {author} {\bibfnamefont {R.~G.}\ \bibnamefont {Carlberg}},\ }\href {\doibase 10.1086/170451} {\bibfield  {journal} {\bibinfo  {journal} {Astrophys. J.}\ }\textbf {\bibinfo {volume} {378}},\ \bibinfo {pages} {496} (\bibinfo {year} {1991})}\BibitemShut {NoStop}%
\bibitem [{\citenamefont {Navarro}\ \emph {et~al.}(1996)\citenamefont {Navarro}, \citenamefont {Frenk},\ and\ \citenamefont {White}}]{Navarro:1995iw}%
  \BibitemOpen
  \bibfield  {author} {\bibinfo {author} {\bibfnamefont {J.~F.}\ \bibnamefont {Navarro}}, \bibinfo {author} {\bibfnamefont {C.~S.}\ \bibnamefont {Frenk}}, \ and\ \bibinfo {author} {\bibfnamefont {S.~D.~M.}\ \bibnamefont {White}},\ }\href {\doibase 10.1086/177173} {\bibfield  {journal} {\bibinfo  {journal} {Astrophys. J.}\ }\textbf {\bibinfo {volume} {462}},\ \bibinfo {pages} {563} (\bibinfo {year} {1996})},\ \Eprint {http://arxiv.org/abs/astro-ph/9508025} {arXiv:astro-ph/9508025} \BibitemShut {NoStop}%
\bibitem [{\citenamefont {Navarro}\ \emph {et~al.}(1997)\citenamefont {Navarro}, \citenamefont {Frenk},\ and\ \citenamefont {White}}]{Navarro:1996gj}%
  \BibitemOpen
  \bibfield  {author} {\bibinfo {author} {\bibfnamefont {J.~F.}\ \bibnamefont {Navarro}}, \bibinfo {author} {\bibfnamefont {C.~S.}\ \bibnamefont {Frenk}}, \ and\ \bibinfo {author} {\bibfnamefont {S.~D.~M.}\ \bibnamefont {White}},\ }\href {\doibase 10.1086/304888} {\bibfield  {journal} {\bibinfo  {journal} {Astrophys. J.}\ }\textbf {\bibinfo {volume} {490}},\ \bibinfo {pages} {493} (\bibinfo {year} {1997})},\ \Eprint {http://arxiv.org/abs/astro-ph/9611107} {arXiv:astro-ph/9611107} \BibitemShut {NoStop}%
\bibitem [{\citenamefont {Martizzi}\ \emph {et~al.}(2012)\citenamefont {Martizzi}, \citenamefont {Teyssier}, \citenamefont {Moore},\ and\ \citenamefont {Wentz}}]{Martizzi:2011aa}%
  \BibitemOpen
  \bibfield  {author} {\bibinfo {author} {\bibfnamefont {D.}~\bibnamefont {Martizzi}}, \bibinfo {author} {\bibfnamefont {R.}~\bibnamefont {Teyssier}}, \bibinfo {author} {\bibfnamefont {B.}~\bibnamefont {Moore}}, \ and\ \bibinfo {author} {\bibfnamefont {T.}~\bibnamefont {Wentz}},\ }\href {\doibase 10.1111/j.1365-2966.2012.20879.x} {\bibfield  {journal} {\bibinfo  {journal} {Mon. Not. Roy. Astron. Soc.}\ }\textbf {\bibinfo {volume} {422}},\ \bibinfo {pages} {3081} (\bibinfo {year} {2012})},\ \Eprint {http://arxiv.org/abs/1112.2752} {arXiv:1112.2752 [astro-ph.CO]} \BibitemShut {NoStop}%
\bibitem [{\citenamefont {Governato}\ \emph {et~al.}(2012)\citenamefont {Governato}, \citenamefont {Zolotov}, \citenamefont {Pontzen}, \citenamefont {Christensen}, \citenamefont {Oh}, \citenamefont {Brooks}, \citenamefont {Quinn}, \citenamefont {Shen},\ and\ \citenamefont {Wadsley}}]{Governato:2012fa}%
  \BibitemOpen
  \bibfield  {author} {\bibinfo {author} {\bibfnamefont {F.}~\bibnamefont {Governato}}, \bibinfo {author} {\bibfnamefont {A.}~\bibnamefont {Zolotov}}, \bibinfo {author} {\bibfnamefont {A.}~\bibnamefont {Pontzen}}, \bibinfo {author} {\bibfnamefont {C.}~\bibnamefont {Christensen}}, \bibinfo {author} {\bibfnamefont {S.~H.}\ \bibnamefont {Oh}}, \bibinfo {author} {\bibfnamefont {A.~M.}\ \bibnamefont {Brooks}}, \bibinfo {author} {\bibfnamefont {T.}~\bibnamefont {Quinn}}, \bibinfo {author} {\bibfnamefont {S.}~\bibnamefont {Shen}}, \ and\ \bibinfo {author} {\bibfnamefont {J.}~\bibnamefont {Wadsley}},\ }\href {\doibase 10.1111/j.1365-2966.2012.20696.x} {\bibfield  {journal} {\bibinfo  {journal} {Mon. Not. Roy. Astron. Soc.}\ }\textbf {\bibinfo {volume} {422}},\ \bibinfo {pages} {1231} (\bibinfo {year} {2012})},\ \Eprint {http://arxiv.org/abs/1202.0554} {arXiv:1202.0554 [astro-ph.CO]} \BibitemShut {NoStop}%
\bibitem [{\citenamefont {Ben\'\i{}tez-Llambay}\ \emph {et~al.}(2019)\citenamefont {Ben\'\i{}tez-Llambay}, \citenamefont {Frenk}, \citenamefont {Ludlow},\ and\ \citenamefont {Navarro}}]{Benitez-Llambay:2019wfi}%
  \BibitemOpen
  \bibfield  {author} {\bibinfo {author} {\bibfnamefont {A.}~\bibnamefont {Ben\'\i{}tez-Llambay}}, \bibinfo {author} {\bibfnamefont {C.~S.}\ \bibnamefont {Frenk}}, \bibinfo {author} {\bibfnamefont {A.~D.}\ \bibnamefont {Ludlow}}, \ and\ \bibinfo {author} {\bibfnamefont {J.~F.}\ \bibnamefont {Navarro}},\ }\href {\doibase 10.1093/mnras/stz1890} {\bibfield  {journal} {\bibinfo  {journal} {Mon. Not. Roy. Astron. Soc.}\ }\textbf {\bibinfo {volume} {488}},\ \bibinfo {pages} {2387} (\bibinfo {year} {2019})},\ \Eprint {http://arxiv.org/abs/1810.04186} {arXiv:1810.04186} \BibitemShut {NoStop}%
\bibitem [{\citenamefont {Spergel}\ and\ \citenamefont {Steinhardt}(2000)}]{Spergel:1999mh}%
  \BibitemOpen
  \bibfield  {author} {\bibinfo {author} {\bibfnamefont {D.~N.}\ \bibnamefont {Spergel}}\ and\ \bibinfo {author} {\bibfnamefont {P.~J.}\ \bibnamefont {Steinhardt}},\ }\href {\doibase 10.1103/PhysRevLett.84.3760} {\bibfield  {journal} {\bibinfo  {journal} {Phys. Rev. Lett.}\ }\textbf {\bibinfo {volume} {84}},\ \bibinfo {pages} {3760} (\bibinfo {year} {2000})},\ \Eprint {http://arxiv.org/abs/astro-ph/9909386} {arXiv:astro-ph/9909386} \BibitemShut {NoStop}%
\bibitem [{\citenamefont {Dave}\ \emph {et~al.}(2001)\citenamefont {Dave}, \citenamefont {Spergel}, \citenamefont {Steinhardt},\ and\ \citenamefont {Wandelt}}]{Dave:2000ar}%
  \BibitemOpen
  \bibfield  {author} {\bibinfo {author} {\bibfnamefont {R.}~\bibnamefont {Dave}}, \bibinfo {author} {\bibfnamefont {D.~N.}\ \bibnamefont {Spergel}}, \bibinfo {author} {\bibfnamefont {P.~J.}\ \bibnamefont {Steinhardt}}, \ and\ \bibinfo {author} {\bibfnamefont {B.~D.}\ \bibnamefont {Wandelt}},\ }\href {\doibase 10.1086/318417} {\bibfield  {journal} {\bibinfo  {journal} {Astrophys. J.}\ }\textbf {\bibinfo {volume} {547}},\ \bibinfo {pages} {574} (\bibinfo {year} {2001})},\ \Eprint {http://arxiv.org/abs/astro-ph/0006218} {arXiv:astro-ph/0006218} \BibitemShut {NoStop}%
\bibitem [{\citenamefont {Vogelsberger}\ \emph {et~al.}(2012)\citenamefont {Vogelsberger}, \citenamefont {Zavala},\ and\ \citenamefont {Loeb}}]{Vogelsberger:2012ku}%
  \BibitemOpen
  \bibfield  {author} {\bibinfo {author} {\bibfnamefont {M.}~\bibnamefont {Vogelsberger}}, \bibinfo {author} {\bibfnamefont {J.}~\bibnamefont {Zavala}}, \ and\ \bibinfo {author} {\bibfnamefont {A.}~\bibnamefont {Loeb}},\ }\href {\doibase 10.1111/j.1365-2966.2012.21182.x} {\bibfield  {journal} {\bibinfo  {journal} {Mon. Not. Roy. Astron. Soc.}\ }\textbf {\bibinfo {volume} {423}},\ \bibinfo {pages} {3740} (\bibinfo {year} {2012})},\ \Eprint {http://arxiv.org/abs/1201.5892} {arXiv:1201.5892 [astro-ph.CO]} \BibitemShut {NoStop}%
\bibitem [{\citenamefont {Rocha}\ \emph {et~al.}(2013)\citenamefont {Rocha}, \citenamefont {Peter}, \citenamefont {Bullock}, \citenamefont {Kaplinghat}, \citenamefont {Garrison-Kimmel}, \citenamefont {Onorbe},\ and\ \citenamefont {Moustakas}}]{Rocha:2012jg}%
  \BibitemOpen
  \bibfield  {author} {\bibinfo {author} {\bibfnamefont {M.}~\bibnamefont {Rocha}}, \bibinfo {author} {\bibfnamefont {A.~H.~G.}\ \bibnamefont {Peter}}, \bibinfo {author} {\bibfnamefont {J.~S.}\ \bibnamefont {Bullock}}, \bibinfo {author} {\bibfnamefont {M.}~\bibnamefont {Kaplinghat}}, \bibinfo {author} {\bibfnamefont {S.}~\bibnamefont {Garrison-Kimmel}}, \bibinfo {author} {\bibfnamefont {J.}~\bibnamefont {Onorbe}}, \ and\ \bibinfo {author} {\bibfnamefont {L.~A.}\ \bibnamefont {Moustakas}},\ }\href {\doibase 10.1093/mnras/sts514} {\bibfield  {journal} {\bibinfo  {journal} {Mon. Not. Roy. Astron. Soc.}\ }\textbf {\bibinfo {volume} {430}},\ \bibinfo {pages} {81} (\bibinfo {year} {2013})},\ \Eprint {http://arxiv.org/abs/1208.3025} {arXiv:1208.3025 [astro-ph.CO]} \BibitemShut {NoStop}%
\bibitem [{\citenamefont {Peter}\ \emph {et~al.}(2013)\citenamefont {Peter}, \citenamefont {Rocha}, \citenamefont {Bullock},\ and\ \citenamefont {Kaplinghat}}]{Peter:2012jh}%
  \BibitemOpen
  \bibfield  {author} {\bibinfo {author} {\bibfnamefont {A.~H.~G.}\ \bibnamefont {Peter}}, \bibinfo {author} {\bibfnamefont {M.}~\bibnamefont {Rocha}}, \bibinfo {author} {\bibfnamefont {J.~S.}\ \bibnamefont {Bullock}}, \ and\ \bibinfo {author} {\bibfnamefont {M.}~\bibnamefont {Kaplinghat}},\ }\href {\doibase 10.1093/mnras/sts535} {\bibfield  {journal} {\bibinfo  {journal} {Mon. Not. Roy. Astron. Soc.}\ }\textbf {\bibinfo {volume} {430}},\ \bibinfo {pages} {105} (\bibinfo {year} {2013})},\ \Eprint {http://arxiv.org/abs/1208.3026} {arXiv:1208.3026 [astro-ph.CO]} \BibitemShut {NoStop}%
\bibitem [{\citenamefont {Elbert}\ \emph {et~al.}(2015)\citenamefont {Elbert}, \citenamefont {Bullock}, \citenamefont {Garrison-Kimmel}, \citenamefont {Rocha}, \citenamefont {O\~norbe},\ and\ \citenamefont {Peter}}]{Elbert:2014bma}%
  \BibitemOpen
  \bibfield  {author} {\bibinfo {author} {\bibfnamefont {O.~D.}\ \bibnamefont {Elbert}}, \bibinfo {author} {\bibfnamefont {J.~S.}\ \bibnamefont {Bullock}}, \bibinfo {author} {\bibfnamefont {S.}~\bibnamefont {Garrison-Kimmel}}, \bibinfo {author} {\bibfnamefont {M.}~\bibnamefont {Rocha}}, \bibinfo {author} {\bibfnamefont {J.}~\bibnamefont {O\~norbe}}, \ and\ \bibinfo {author} {\bibfnamefont {A.~H.~G.}\ \bibnamefont {Peter}},\ }\href {\doibase 10.1093/mnras/stv1470} {\bibfield  {journal} {\bibinfo  {journal} {Mon. Not. Roy. Astron. Soc.}\ }\textbf {\bibinfo {volume} {453}},\ \bibinfo {pages} {29} (\bibinfo {year} {2015})},\ \Eprint {http://arxiv.org/abs/1412.1477} {arXiv:1412.1477 [astro-ph.GA]} \BibitemShut {NoStop}%
\bibitem [{\citenamefont {Kaplinghat}\ \emph {et~al.}(2016)\citenamefont {Kaplinghat}, \citenamefont {Tulin},\ and\ \citenamefont {Yu}}]{Kaplinghat:2015aga}%
  \BibitemOpen
  \bibfield  {author} {\bibinfo {author} {\bibfnamefont {M.}~\bibnamefont {Kaplinghat}}, \bibinfo {author} {\bibfnamefont {S.}~\bibnamefont {Tulin}}, \ and\ \bibinfo {author} {\bibfnamefont {H.-B.}\ \bibnamefont {Yu}},\ }\href {\doibase 10.1103/PhysRevLett.116.041302} {\bibfield  {journal} {\bibinfo  {journal} {Phys. Rev. Lett.}\ }\textbf {\bibinfo {volume} {116}},\ \bibinfo {pages} {041302} (\bibinfo {year} {2016})},\ \Eprint {http://arxiv.org/abs/1508.03339} {arXiv:1508.03339 [astro-ph.CO]} \BibitemShut {NoStop}%
\bibitem [{\citenamefont {Harigaya}\ and\ \citenamefont {Nomura}(2016)}]{Harigaya:2016rwr}%
  \BibitemOpen
  \bibfield  {author} {\bibinfo {author} {\bibfnamefont {K.}~\bibnamefont {Harigaya}}\ and\ \bibinfo {author} {\bibfnamefont {Y.}~\bibnamefont {Nomura}},\ }\href {\doibase 10.1103/PhysRevD.94.035013} {\bibfield  {journal} {\bibinfo  {journal} {Phys. Rev. D}\ }\textbf {\bibinfo {volume} {94}},\ \bibinfo {pages} {035013} (\bibinfo {year} {2016})},\ \Eprint {http://arxiv.org/abs/1603.03430} {arXiv:1603.03430 [hep-ph]} \BibitemShut {NoStop}%
\bibitem [{\citenamefont {Co}\ \emph {et~al.}(2017)\citenamefont {Co}, \citenamefont {Harigaya},\ and\ \citenamefont {Nomura}}]{Co:2016akw}%
  \BibitemOpen
  \bibfield  {author} {\bibinfo {author} {\bibfnamefont {R.~T.}\ \bibnamefont {Co}}, \bibinfo {author} {\bibfnamefont {K.}~\bibnamefont {Harigaya}}, \ and\ \bibinfo {author} {\bibfnamefont {Y.}~\bibnamefont {Nomura}},\ }\href {\doibase 10.1103/PhysRevLett.118.101801} {\bibfield  {journal} {\bibinfo  {journal} {Phys. Rev. Lett.}\ }\textbf {\bibinfo {volume} {118}},\ \bibinfo {pages} {101801} (\bibinfo {year} {2017})},\ \Eprint {http://arxiv.org/abs/1610.03848} {arXiv:1610.03848 [hep-ph]} \BibitemShut {NoStop}%
\bibitem [{\citenamefont {Ibe}\ \emph {et~al.}(2021)\citenamefont {Ibe}, \citenamefont {Kobayashi},\ and\ \citenamefont {Watanabe}}]{Ibe:2021gil}%
  \BibitemOpen
  \bibfield  {author} {\bibinfo {author} {\bibfnamefont {M.}~\bibnamefont {Ibe}}, \bibinfo {author} {\bibfnamefont {S.}~\bibnamefont {Kobayashi}}, \ and\ \bibinfo {author} {\bibfnamefont {K.}~\bibnamefont {Watanabe}},\ }\href {\doibase 10.1007/JHEP07(2021)220} {\bibfield  {journal} {\bibinfo  {journal} {JHEP}\ }\textbf {\bibinfo {volume} {07}},\ \bibinfo {pages} {220} (\bibinfo {year} {2021})},\ \Eprint {http://arxiv.org/abs/2105.07642} {arXiv:2105.07642 [hep-ph]} \BibitemShut {NoStop}%
\bibitem [{\citenamefont {Nambu}\ and\ \citenamefont {Jona-Lasinio}(1961{\natexlab{a}})}]{Nambu:1961tp}%
  \BibitemOpen
  \bibfield  {author} {\bibinfo {author} {\bibfnamefont {Y.}~\bibnamefont {Nambu}}\ and\ \bibinfo {author} {\bibfnamefont {G.}~\bibnamefont {Jona-Lasinio}},\ }\href {\doibase 10.1103/PhysRev.122.345} {\bibfield  {journal} {\bibinfo  {journal} {Phys. Rev.}\ }\textbf {\bibinfo {volume} {122}},\ \bibinfo {pages} {345} (\bibinfo {year} {1961}{\natexlab{a}})}\BibitemShut {NoStop}%
\bibitem [{\citenamefont {Nambu}\ and\ \citenamefont {Jona-Lasinio}(1961{\natexlab{b}})}]{Nambu:1961fr}%
  \BibitemOpen
  \bibfield  {author} {\bibinfo {author} {\bibfnamefont {Y.}~\bibnamefont {Nambu}}\ and\ \bibinfo {author} {\bibfnamefont {G.}~\bibnamefont {Jona-Lasinio}},\ }\href {\doibase 10.1103/PhysRev.124.246} {\bibfield  {journal} {\bibinfo  {journal} {Phys. Rev.}\ }\textbf {\bibinfo {volume} {124}},\ \bibinfo {pages} {246} (\bibinfo {year} {1961}{\natexlab{b}})}\BibitemShut {NoStop}%
\bibitem [{\citenamefont {Klevansky}(1992)}]{Klevansky:1992qe}%
  \BibitemOpen
  \bibfield  {author} {\bibinfo {author} {\bibfnamefont {S.~P.}\ \bibnamefont {Klevansky}},\ }\href {\doibase 10.1103/RevModPhys.64.649} {\bibfield  {journal} {\bibinfo  {journal} {Rev. Mod. Phys.}\ }\textbf {\bibinfo {volume} {64}},\ \bibinfo {pages} {649} (\bibinfo {year} {1992})}\BibitemShut {NoStop}%
\bibitem [{\citenamefont {Hatsuda}\ and\ \citenamefont {Kunihiro}(1994)}]{Hatsuda:1994pi}%
  \BibitemOpen
  \bibfield  {author} {\bibinfo {author} {\bibfnamefont {T.}~\bibnamefont {Hatsuda}}\ and\ \bibinfo {author} {\bibfnamefont {T.}~\bibnamefont {Kunihiro}},\ }\href {\doibase 10.1016/0370-1573(94)90022-1} {\bibfield  {journal} {\bibinfo  {journal} {Phys. Rept.}\ }\textbf {\bibinfo {volume} {247}},\ \bibinfo {pages} {221} (\bibinfo {year} {1994})},\ \Eprint {http://arxiv.org/abs/hep-ph/9401310} {arXiv:hep-ph/9401310} \BibitemShut {NoStop}%
\bibitem [{\citenamefont {Antipin}\ \emph {et~al.}(2015)\citenamefont {Antipin}, \citenamefont {Redi}, \citenamefont {Strumia},\ and\ \citenamefont {Vigiani}}]{Antipin:2015xia}%
  \BibitemOpen
  \bibfield  {author} {\bibinfo {author} {\bibfnamefont {O.}~\bibnamefont {Antipin}}, \bibinfo {author} {\bibfnamefont {M.}~\bibnamefont {Redi}}, \bibinfo {author} {\bibfnamefont {A.}~\bibnamefont {Strumia}}, \ and\ \bibinfo {author} {\bibfnamefont {E.}~\bibnamefont {Vigiani}},\ }\href {\doibase 10.1007/JHEP07(2015)039} {\bibfield  {journal} {\bibinfo  {journal} {JHEP}\ }\textbf {\bibinfo {volume} {07}},\ \bibinfo {pages} {039} (\bibinfo {year} {2015})},\ \Eprint {http://arxiv.org/abs/1503.08749} {arXiv:1503.08749 [hep-ph]} \BibitemShut {NoStop}%
\bibitem [{\citenamefont {Morrison}\ \emph {et~al.}(2021)\citenamefont {Morrison}, \citenamefont {Profumo},\ and\ \citenamefont {Robinson}}]{Morrison:2020yeg}%
  \BibitemOpen
  \bibfield  {author} {\bibinfo {author} {\bibfnamefont {L.}~\bibnamefont {Morrison}}, \bibinfo {author} {\bibfnamefont {S.}~\bibnamefont {Profumo}}, \ and\ \bibinfo {author} {\bibfnamefont {D.~J.}\ \bibnamefont {Robinson}},\ }\href {\doibase 10.1088/1475-7516/2021/05/058} {\bibfield  {journal} {\bibinfo  {journal} {JCAP}\ }\textbf {\bibinfo {volume} {05}},\ \bibinfo {pages} {058} (\bibinfo {year} {2021})},\ \Eprint {http://arxiv.org/abs/2010.03586} {arXiv:2010.03586 [hep-ph]} \BibitemShut {NoStop}%
\bibitem [{\citenamefont {Garani}\ \emph {et~al.}(2021)\citenamefont {Garani}, \citenamefont {Redi},\ and\ \citenamefont {Tesi}}]{Garani:2021zrr}%
  \BibitemOpen
  \bibfield  {author} {\bibinfo {author} {\bibfnamefont {R.}~\bibnamefont {Garani}}, \bibinfo {author} {\bibfnamefont {M.}~\bibnamefont {Redi}}, \ and\ \bibinfo {author} {\bibfnamefont {A.}~\bibnamefont {Tesi}},\ }\href {\doibase 10.1007/JHEP12(2021)139} {\bibfield  {journal} {\bibinfo  {journal} {JHEP}\ }\textbf {\bibinfo {volume} {12}},\ \bibinfo {pages} {139} (\bibinfo {year} {2021})},\ \Eprint {http://arxiv.org/abs/2105.03429} {arXiv:2105.03429 [hep-ph]} \BibitemShut {NoStop}%
\bibitem [{\citenamefont {Holdom}(1986)}]{Holdom:1985ag}%
  \BibitemOpen
  \bibfield  {author} {\bibinfo {author} {\bibfnamefont {B.}~\bibnamefont {Holdom}},\ }\href {\doibase 10.1016/0370-2693(86)91377-8} {\bibfield  {journal} {\bibinfo  {journal} {Phys. Lett. B}\ }\textbf {\bibinfo {volume} {166}},\ \bibinfo {pages} {196} (\bibinfo {year} {1986})}\BibitemShut {NoStop}%
\bibitem [{\citenamefont {Andrade}\ \emph {et~al.}(2021)\citenamefont {Andrade}, \citenamefont {Fuson}, \citenamefont {Gad-Nasr}, \citenamefont {Kong}, \citenamefont {Minor}, \citenamefont {Roberts},\ and\ \citenamefont {Kaplinghat}}]{Andrade:2020lqq}%
  \BibitemOpen
  \bibfield  {author} {\bibinfo {author} {\bibfnamefont {K.~E.}\ \bibnamefont {Andrade}}, \bibinfo {author} {\bibfnamefont {J.}~\bibnamefont {Fuson}}, \bibinfo {author} {\bibfnamefont {S.}~\bibnamefont {Gad-Nasr}}, \bibinfo {author} {\bibfnamefont {D.}~\bibnamefont {Kong}}, \bibinfo {author} {\bibfnamefont {Q.}~\bibnamefont {Minor}}, \bibinfo {author} {\bibfnamefont {M.~G.}\ \bibnamefont {Roberts}}, \ and\ \bibinfo {author} {\bibfnamefont {M.}~\bibnamefont {Kaplinghat}},\ }\href {\doibase 10.1093/mnras/stab3241} {\bibfield  {journal} {\bibinfo  {journal} {Mon. Not. Roy. Astron. Soc.}\ }\textbf {\bibinfo {volume} {510}},\ \bibinfo {pages} {54} (\bibinfo {year} {2021})},\ \Eprint {http://arxiv.org/abs/2012.06611} {arXiv:2012.06611 [astro-ph.CO]} \BibitemShut {NoStop}%
\bibitem [{\citenamefont {Feng}\ \emph {et~al.}(2010)\citenamefont {Feng}, \citenamefont {Kaplinghat},\ and\ \citenamefont {Yu}}]{Feng:2009hw}%
  \BibitemOpen
  \bibfield  {author} {\bibinfo {author} {\bibfnamefont {J.~L.}\ \bibnamefont {Feng}}, \bibinfo {author} {\bibfnamefont {M.}~\bibnamefont {Kaplinghat}}, \ and\ \bibinfo {author} {\bibfnamefont {H.-B.}\ \bibnamefont {Yu}},\ }\href {\doibase 10.1103/PhysRevLett.104.151301} {\bibfield  {journal} {\bibinfo  {journal} {Phys. Rev. Lett.}\ }\textbf {\bibinfo {volume} {104}},\ \bibinfo {pages} {151301} (\bibinfo {year} {2010})},\ \Eprint {http://arxiv.org/abs/0911.0422} {arXiv:0911.0422 [hep-ph]} \BibitemShut {NoStop}%
\bibitem [{\citenamefont {Ibe}\ \emph {et~al.}(2020)\citenamefont {Ibe}, \citenamefont {Kobayashi}, \citenamefont {Nakayama},\ and\ \citenamefont {Shirai}}]{Ibe:2019gpv}%
  \BibitemOpen
  \bibfield  {author} {\bibinfo {author} {\bibfnamefont {M.}~\bibnamefont {Ibe}}, \bibinfo {author} {\bibfnamefont {S.}~\bibnamefont {Kobayashi}}, \bibinfo {author} {\bibfnamefont {Y.}~\bibnamefont {Nakayama}}, \ and\ \bibinfo {author} {\bibfnamefont {S.}~\bibnamefont {Shirai}},\ }\href {\doibase 10.1007/JHEP04(2020)009} {\bibfield  {journal} {\bibinfo  {journal} {JHEP}\ }\textbf {\bibinfo {volume} {04}},\ \bibinfo {pages} {009} (\bibinfo {year} {2020})},\ \Eprint {http://arxiv.org/abs/1912.12152} {arXiv:1912.12152 [hep-ph]} \BibitemShut {NoStop}%
\bibitem [{\citenamefont {Li}\ \emph {et~al.}(2020)\citenamefont {Li}, \citenamefont {Fuller},\ and\ \citenamefont {Grohs}}]{Li:2020roy}%
  \BibitemOpen
  \bibfield  {author} {\bibinfo {author} {\bibfnamefont {J.-T.}\ \bibnamefont {Li}}, \bibinfo {author} {\bibfnamefont {G.~M.}\ \bibnamefont {Fuller}}, \ and\ \bibinfo {author} {\bibfnamefont {E.}~\bibnamefont {Grohs}},\ }\href {\doibase 10.1088/1475-7516/2020/12/049} {\bibfield  {journal} {\bibinfo  {journal} {JCAP}\ }\textbf {\bibinfo {volume} {12}},\ \bibinfo {pages} {049} (\bibinfo {year} {2020})},\ \Eprint {http://arxiv.org/abs/2009.14325} {arXiv:2009.14325 [astro-ph.CO]} \BibitemShut {NoStop}%
\bibitem [{\citenamefont {Calabrese}\ \emph {et~al.}(2025)\citenamefont {Calabrese} \emph {et~al.}}]{ACT:2025tim}%
  \BibitemOpen
  \bibfield  {author} {\bibinfo {author} {\bibfnamefont {E.}~\bibnamefont {Calabrese}} \emph {et~al.} (\bibinfo {collaboration} {ACT}),\ }\href@noop {} {\  (\bibinfo {year} {2025})},\ \Eprint {http://arxiv.org/abs/2503.14454} {arXiv:2503.14454 [astro-ph.CO]} \BibitemShut {NoStop}%
\bibitem [{\citenamefont {Akita}\ and\ \citenamefont {Yamaguchi}(2020)}]{Akita:2020szl}%
  \BibitemOpen
  \bibfield  {author} {\bibinfo {author} {\bibfnamefont {K.}~\bibnamefont {Akita}}\ and\ \bibinfo {author} {\bibfnamefont {M.}~\bibnamefont {Yamaguchi}},\ }\href {\doibase 10.1088/1475-7516/2020/08/012} {\bibfield  {journal} {\bibinfo  {journal} {JCAP}\ }\textbf {\bibinfo {volume} {08}},\ \bibinfo {pages} {012} (\bibinfo {year} {2020})},\ \Eprint {http://arxiv.org/abs/2005.07047} {arXiv:2005.07047 [hep-ph]} \BibitemShut {NoStop}%
\bibitem [{\citenamefont {Froustey}\ \emph {et~al.}(2020)\citenamefont {Froustey}, \citenamefont {Pitrou},\ and\ \citenamefont {Volpe}}]{Froustey:2020mcq}%
  \BibitemOpen
  \bibfield  {author} {\bibinfo {author} {\bibfnamefont {J.}~\bibnamefont {Froustey}}, \bibinfo {author} {\bibfnamefont {C.}~\bibnamefont {Pitrou}}, \ and\ \bibinfo {author} {\bibfnamefont {M.~C.}\ \bibnamefont {Volpe}},\ }\href {\doibase 10.1088/1475-7516/2020/12/015} {\bibfield  {journal} {\bibinfo  {journal} {JCAP}\ }\textbf {\bibinfo {volume} {12}},\ \bibinfo {pages} {015} (\bibinfo {year} {2020})},\ \Eprint {http://arxiv.org/abs/2008.01074} {arXiv:2008.01074 [hep-ph]} \BibitemShut {NoStop}%
\bibitem [{\citenamefont {Bennett}\ \emph {et~al.}(2021)\citenamefont {Bennett}, \citenamefont {Buldgen}, \citenamefont {De~Salas}, \citenamefont {Drewes}, \citenamefont {Gariazzo}, \citenamefont {Pastor},\ and\ \citenamefont {Wong}}]{Bennett:2020zkv}%
  \BibitemOpen
  \bibfield  {author} {\bibinfo {author} {\bibfnamefont {J.~J.}\ \bibnamefont {Bennett}}, \bibinfo {author} {\bibfnamefont {G.}~\bibnamefont {Buldgen}}, \bibinfo {author} {\bibfnamefont {P.~F.}\ \bibnamefont {De~Salas}}, \bibinfo {author} {\bibfnamefont {M.}~\bibnamefont {Drewes}}, \bibinfo {author} {\bibfnamefont {S.}~\bibnamefont {Gariazzo}}, \bibinfo {author} {\bibfnamefont {S.}~\bibnamefont {Pastor}}, \ and\ \bibinfo {author} {\bibfnamefont {Y.~Y.~Y.}\ \bibnamefont {Wong}},\ }\href {\doibase 10.1088/1475-7516/2021/04/073} {\bibfield  {journal} {\bibinfo  {journal} {JCAP}\ }\textbf {\bibinfo {volume} {04}},\ \bibinfo {pages} {073} (\bibinfo {year} {2021})},\ \Eprint {http://arxiv.org/abs/2012.02726} {arXiv:2012.02726 [hep-ph]} \BibitemShut {NoStop}%
\bibitem [{\citenamefont {Drewes}\ \emph {et~al.}(2024)\citenamefont {Drewes}, \citenamefont {Georis}, \citenamefont {Klasen}, \citenamefont {Wiggering},\ and\ \citenamefont {Wong}}]{Drewes:2024wbw}%
  \BibitemOpen
  \bibfield  {author} {\bibinfo {author} {\bibfnamefont {M.}~\bibnamefont {Drewes}}, \bibinfo {author} {\bibfnamefont {Y.}~\bibnamefont {Georis}}, \bibinfo {author} {\bibfnamefont {M.}~\bibnamefont {Klasen}}, \bibinfo {author} {\bibfnamefont {L.~P.}\ \bibnamefont {Wiggering}}, \ and\ \bibinfo {author} {\bibfnamefont {Y.~Y.~Y.}\ \bibnamefont {Wong}},\ }\href {\doibase 10.1088/1475-7516/2024/06/032} {\bibfield  {journal} {\bibinfo  {journal} {JCAP}\ }\textbf {\bibinfo {volume} {06}},\ \bibinfo {pages} {032} (\bibinfo {year} {2024})},\ \Eprint {http://arxiv.org/abs/2402.18481} {arXiv:2402.18481 [hep-ph]} \BibitemShut {NoStop}%
\bibitem [{\citenamefont {Fornengo}\ \emph {et~al.}(2011)\citenamefont {Fornengo}, \citenamefont {Panci},\ and\ \citenamefont {Regis}}]{Fornengo:2011sz}%
  \BibitemOpen
  \bibfield  {author} {\bibinfo {author} {\bibfnamefont {N.}~\bibnamefont {Fornengo}}, \bibinfo {author} {\bibfnamefont {P.}~\bibnamefont {Panci}}, \ and\ \bibinfo {author} {\bibfnamefont {M.}~\bibnamefont {Regis}},\ }\href {\doibase 10.1103/PhysRevD.84.115002} {\bibfield  {journal} {\bibinfo  {journal} {Phys. Rev. D}\ }\textbf {\bibinfo {volume} {84}},\ \bibinfo {pages} {115002} (\bibinfo {year} {2011})},\ \Eprint {http://arxiv.org/abs/1108.4661} {arXiv:1108.4661 [hep-ph]} \BibitemShut {NoStop}%
\bibitem [{\citenamefont {Kaplinghat}\ \emph {et~al.}(2014)\citenamefont {Kaplinghat}, \citenamefont {Tulin},\ and\ \citenamefont {Yu}}]{Kaplinghat:2013yxa}%
  \BibitemOpen
  \bibfield  {author} {\bibinfo {author} {\bibfnamefont {M.}~\bibnamefont {Kaplinghat}}, \bibinfo {author} {\bibfnamefont {S.}~\bibnamefont {Tulin}}, \ and\ \bibinfo {author} {\bibfnamefont {H.-B.}\ \bibnamefont {Yu}},\ }\href {\doibase 10.1103/PhysRevD.89.035009} {\bibfield  {journal} {\bibinfo  {journal} {Phys. Rev. D}\ }\textbf {\bibinfo {volume} {89}},\ \bibinfo {pages} {035009} (\bibinfo {year} {2014})},\ \Eprint {http://arxiv.org/abs/1310.7945} {arXiv:1310.7945 [hep-ph]} \BibitemShut {NoStop}%
\bibitem [{\citenamefont {Del~Nobile}\ \emph {et~al.}(2015)\citenamefont {Del~Nobile}, \citenamefont {Kaplinghat},\ and\ \citenamefont {Yu}}]{DelNobile:2015uua}%
  \BibitemOpen
  \bibfield  {author} {\bibinfo {author} {\bibfnamefont {E.}~\bibnamefont {Del~Nobile}}, \bibinfo {author} {\bibfnamefont {M.}~\bibnamefont {Kaplinghat}}, \ and\ \bibinfo {author} {\bibfnamefont {H.-B.}\ \bibnamefont {Yu}},\ }\href {\doibase 10.1088/1475-7516/2015/10/055} {\bibfield  {journal} {\bibinfo  {journal} {JCAP}\ }\textbf {\bibinfo {volume} {10}},\ \bibinfo {pages} {055} (\bibinfo {year} {2015})},\ \Eprint {http://arxiv.org/abs/1507.04007} {arXiv:1507.04007 [hep-ph]} \BibitemShut {NoStop}%
\bibitem [{\citenamefont {Ren}\ \emph {et~al.}(2018)\citenamefont {Ren} \emph {et~al.}}]{PandaX-II:2018xpz}%
  \BibitemOpen
  \bibfield  {author} {\bibinfo {author} {\bibfnamefont {X.}~\bibnamefont {Ren}} \emph {et~al.} (\bibinfo {collaboration} {PandaX-II}),\ }\href {\doibase 10.1103/PhysRevLett.121.021304} {\bibfield  {journal} {\bibinfo  {journal} {Phys. Rev. Lett.}\ }\textbf {\bibinfo {volume} {121}},\ \bibinfo {pages} {021304} (\bibinfo {year} {2018})},\ \Eprint {http://arxiv.org/abs/1802.06912} {arXiv:1802.06912 [hep-ph]} \BibitemShut {NoStop}%
\bibitem [{\citenamefont {Aprile}\ \emph {et~al.}(2019)\citenamefont {Aprile} \emph {et~al.}}]{XENON:2019gfn}%
  \BibitemOpen
  \bibfield  {author} {\bibinfo {author} {\bibfnamefont {E.}~\bibnamefont {Aprile}} \emph {et~al.} (\bibinfo {collaboration} {XENON}),\ }\href {\doibase 10.1103/PhysRevLett.123.251801} {\bibfield  {journal} {\bibinfo  {journal} {Phys. Rev. Lett.}\ }\textbf {\bibinfo {volume} {123}},\ \bibinfo {pages} {251801} (\bibinfo {year} {2019})},\ \Eprint {http://arxiv.org/abs/1907.11485} {arXiv:1907.11485 [hep-ex]} \BibitemShut {NoStop}%
\bibitem [{\citenamefont {Huang}\ \emph {et~al.}(2023)\citenamefont {Huang} \emph {et~al.}}]{PandaX:2023xgl}%
  \BibitemOpen
  \bibfield  {author} {\bibinfo {author} {\bibfnamefont {D.}~\bibnamefont {Huang}} \emph {et~al.} (\bibinfo {collaboration} {PandaX}),\ }\href {\doibase 10.1103/PhysRevLett.131.191002} {\bibfield  {journal} {\bibinfo  {journal} {Phys. Rev. Lett.}\ }\textbf {\bibinfo {volume} {131}},\ \bibinfo {pages} {191002} (\bibinfo {year} {2023})},\ \Eprint {http://arxiv.org/abs/2308.01540} {arXiv:2308.01540 [hep-ex]} \BibitemShut {NoStop}%
\bibitem [{\citenamefont {Aprile}\ \emph {et~al.}(2025)\citenamefont {Aprile} \emph {et~al.}}]{XENON:2024hup}%
  \BibitemOpen
  \bibfield  {author} {\bibinfo {author} {\bibfnamefont {E.}~\bibnamefont {Aprile}} \emph {et~al.} (\bibinfo {collaboration} {XENON}),\ }\href {\doibase 10.1103/PhysRevLett.134.111802} {\bibfield  {journal} {\bibinfo  {journal} {Phys. Rev. Lett.}\ }\textbf {\bibinfo {volume} {134}},\ \bibinfo {pages} {111802} (\bibinfo {year} {2025})},\ \Eprint {http://arxiv.org/abs/2409.17868} {arXiv:2409.17868 [hep-ex]} \BibitemShut {NoStop}%
\bibitem [{\citenamefont {Chu}\ \emph {et~al.}(2016)\citenamefont {Chu}, \citenamefont {Garcia-Cely},\ and\ \citenamefont {Hambye}}]{Chu:2016pew}%
  \BibitemOpen
  \bibfield  {author} {\bibinfo {author} {\bibfnamefont {X.}~\bibnamefont {Chu}}, \bibinfo {author} {\bibfnamefont {C.}~\bibnamefont {Garcia-Cely}}, \ and\ \bibinfo {author} {\bibfnamefont {T.}~\bibnamefont {Hambye}},\ }\href {\doibase 10.1007/JHEP11(2016)048} {\bibfield  {journal} {\bibinfo  {journal} {JHEP}\ }\textbf {\bibinfo {volume} {11}},\ \bibinfo {pages} {048} (\bibinfo {year} {2016})},\ \Eprint {http://arxiv.org/abs/1609.00399} {arXiv:1609.00399 [hep-ph]} \BibitemShut {NoStop}%
\bibitem [{\citenamefont {Chao}(2019)}]{Chao:2019lhb}%
  \BibitemOpen
  \bibfield  {author} {\bibinfo {author} {\bibfnamefont {W.}~\bibnamefont {Chao}},\ }\href {\doibase 10.1007/JHEP11(2019)013} {\bibfield  {journal} {\bibinfo  {journal} {JHEP}\ }\textbf {\bibinfo {volume} {11}},\ \bibinfo {pages} {013} (\bibinfo {year} {2019})},\ \Eprint {http://arxiv.org/abs/1904.09785} {arXiv:1904.09785 [hep-ph]} \BibitemShut {NoStop}%
\bibitem [{\citenamefont {Agnese}\ \emph {et~al.}(2019)\citenamefont {Agnese} \emph {et~al.}}]{SuperCDMS:2018gro}%
  \BibitemOpen
  \bibfield  {author} {\bibinfo {author} {\bibfnamefont {R.}~\bibnamefont {Agnese}} \emph {et~al.} (\bibinfo {collaboration} {SuperCDMS}),\ }\href {\doibase 10.1103/PhysRevD.99.062001} {\bibfield  {journal} {\bibinfo  {journal} {Phys. Rev. D}\ }\textbf {\bibinfo {volume} {99}},\ \bibinfo {pages} {062001} (\bibinfo {year} {2019})},\ \Eprint {http://arxiv.org/abs/1808.09098} {arXiv:1808.09098 [astro-ph.CO]} \BibitemShut {NoStop}%
\bibitem [{\citenamefont {Abdelhameed}\ \emph {et~al.}(2019)\citenamefont {Abdelhameed} \emph {et~al.}}]{CRESST:2019jnq}%
  \BibitemOpen
  \bibfield  {author} {\bibinfo {author} {\bibfnamefont {A.~H.}\ \bibnamefont {Abdelhameed}} \emph {et~al.} (\bibinfo {collaboration} {CRESST}),\ }\href {\doibase 10.1103/PhysRevD.100.102002} {\bibfield  {journal} {\bibinfo  {journal} {Phys. Rev. D}\ }\textbf {\bibinfo {volume} {100}},\ \bibinfo {pages} {102002} (\bibinfo {year} {2019})},\ \Eprint {http://arxiv.org/abs/1904.00498} {arXiv:1904.00498 [astro-ph.CO]} \BibitemShut {NoStop}%
\bibitem [{\citenamefont {Agnes}\ \emph {et~al.}(2023)\citenamefont {Agnes} \emph {et~al.}}]{GlobalArgonDarkMatter:2022ppc}%
  \BibitemOpen
  \bibfield  {author} {\bibinfo {author} {\bibfnamefont {P.}~\bibnamefont {Agnes}} \emph {et~al.} (\bibinfo {collaboration} {Global Argon Dark Matter}),\ }\href {\doibase 10.1103/PhysRevD.107.112006} {\bibfield  {journal} {\bibinfo  {journal} {Phys. Rev. D}\ }\textbf {\bibinfo {volume} {107}},\ \bibinfo {pages} {112006} (\bibinfo {year} {2023})},\ \Eprint {http://arxiv.org/abs/2209.01177} {arXiv:2209.01177 [physics.ins-det]} \BibitemShut {NoStop}%
\bibitem [{\citenamefont {Acerbi}\ \emph {et~al.}(2024)\citenamefont {Acerbi} \emph {et~al.}}]{DarkSide-20k:2024yfq}%
  \BibitemOpen
  \bibfield  {author} {\bibinfo {author} {\bibfnamefont {F.}~\bibnamefont {Acerbi}} \emph {et~al.} (\bibinfo {collaboration} {DarkSide-20k}),\ }\href {\doibase 10.1038/s42005-024-01896-z} {\bibfield  {journal} {\bibinfo  {journal} {Commun. Phys.}\ }\textbf {\bibinfo {volume} {7}},\ \bibinfo {pages} {422} (\bibinfo {year} {2024})},\ \Eprint {http://arxiv.org/abs/2407.05813} {arXiv:2407.05813 [hep-ex]} \BibitemShut {NoStop}%
\bibitem [{\citenamefont {Bjorken}\ \emph {et~al.}(1988)\citenamefont {Bjorken}, \citenamefont {Ecklund}, \citenamefont {Nelson}, \citenamefont {Abashian}, \citenamefont {Church}, \citenamefont {Lu}, \citenamefont {Mo}, \citenamefont {Nunamaker},\ and\ \citenamefont {Rassmann}}]{Bjorken:1988as}%
  \BibitemOpen
  \bibfield  {author} {\bibinfo {author} {\bibfnamefont {J.~D.}\ \bibnamefont {Bjorken}}, \bibinfo {author} {\bibfnamefont {S.}~\bibnamefont {Ecklund}}, \bibinfo {author} {\bibfnamefont {W.~R.}\ \bibnamefont {Nelson}}, \bibinfo {author} {\bibfnamefont {A.}~\bibnamefont {Abashian}}, \bibinfo {author} {\bibfnamefont {C.}~\bibnamefont {Church}}, \bibinfo {author} {\bibfnamefont {B.}~\bibnamefont {Lu}}, \bibinfo {author} {\bibfnamefont {L.~W.}\ \bibnamefont {Mo}}, \bibinfo {author} {\bibfnamefont {T.~A.}\ \bibnamefont {Nunamaker}}, \ and\ \bibinfo {author} {\bibfnamefont {P.}~\bibnamefont {Rassmann}},\ }\href {\doibase 10.1103/PhysRevD.38.3375} {\bibfield  {journal} {\bibinfo  {journal} {Phys. Rev. D}\ }\textbf {\bibinfo {volume} {38}},\ \bibinfo {pages} {3375} (\bibinfo {year} {1988})}\BibitemShut {NoStop}%
\bibitem [{\citenamefont {Chakraborti}\ \emph {et~al.}(2021)\citenamefont {Chakraborti}, \citenamefont {Feng}, \citenamefont {Koga},\ and\ \citenamefont {Valli}}]{Chakraborti:2021hfm}%
  \BibitemOpen
  \bibfield  {author} {\bibinfo {author} {\bibfnamefont {S.}~\bibnamefont {Chakraborti}}, \bibinfo {author} {\bibfnamefont {J.~L.}\ \bibnamefont {Feng}}, \bibinfo {author} {\bibfnamefont {J.~K.}\ \bibnamefont {Koga}}, \ and\ \bibinfo {author} {\bibfnamefont {M.}~\bibnamefont {Valli}},\ }\href {\doibase 10.1103/PhysRevD.104.055023} {\bibfield  {journal} {\bibinfo  {journal} {Phys. Rev. D}\ }\textbf {\bibinfo {volume} {104}},\ \bibinfo {pages} {055023} (\bibinfo {year} {2021})},\ \Eprint {http://arxiv.org/abs/2105.10289} {arXiv:2105.10289 [hep-ph]} \BibitemShut {NoStop}%
\bibitem [{\citenamefont {Bjorken}\ \emph {et~al.}(2009)\citenamefont {Bjorken}, \citenamefont {Essig}, \citenamefont {Schuster},\ and\ \citenamefont {Toro}}]{Bjorken:2009mm}%
  \BibitemOpen
  \bibfield  {author} {\bibinfo {author} {\bibfnamefont {J.~D.}\ \bibnamefont {Bjorken}}, \bibinfo {author} {\bibfnamefont {R.}~\bibnamefont {Essig}}, \bibinfo {author} {\bibfnamefont {P.}~\bibnamefont {Schuster}}, \ and\ \bibinfo {author} {\bibfnamefont {N.}~\bibnamefont {Toro}},\ }\href {\doibase 10.1103/PhysRevD.80.075018} {\bibfield  {journal} {\bibinfo  {journal} {Phys. Rev. D}\ }\textbf {\bibinfo {volume} {80}},\ \bibinfo {pages} {075018} (\bibinfo {year} {2009})},\ \Eprint {http://arxiv.org/abs/0906.0580} {arXiv:0906.0580 [hep-ph]} \BibitemShut {NoStop}%
\bibitem [{\citenamefont {Andreas}\ \emph {et~al.}(2012)\citenamefont {Andreas}, \citenamefont {Niebuhr},\ and\ \citenamefont {Ringwald}}]{Andreas:2012mt}%
  \BibitemOpen
  \bibfield  {author} {\bibinfo {author} {\bibfnamefont {S.}~\bibnamefont {Andreas}}, \bibinfo {author} {\bibfnamefont {C.}~\bibnamefont {Niebuhr}}, \ and\ \bibinfo {author} {\bibfnamefont {A.}~\bibnamefont {Ringwald}},\ }\href {\doibase 10.1103/PhysRevD.86.095019} {\bibfield  {journal} {\bibinfo  {journal} {Phys. Rev. D}\ }\textbf {\bibinfo {volume} {86}},\ \bibinfo {pages} {095019} (\bibinfo {year} {2012})},\ \Eprint {http://arxiv.org/abs/1209.6083} {arXiv:1209.6083 [hep-ph]} \BibitemShut {NoStop}%
\bibitem [{\citenamefont {Chang}\ \emph {et~al.}(2017)\citenamefont {Chang}, \citenamefont {Essig},\ and\ \citenamefont {McDermott}}]{Chang:2016ntp}%
  \BibitemOpen
  \bibfield  {author} {\bibinfo {author} {\bibfnamefont {J.~H.}\ \bibnamefont {Chang}}, \bibinfo {author} {\bibfnamefont {R.}~\bibnamefont {Essig}}, \ and\ \bibinfo {author} {\bibfnamefont {S.~D.}\ \bibnamefont {McDermott}},\ }\href {\doibase 10.1007/JHEP01(2017)107} {\bibfield  {journal} {\bibinfo  {journal} {JHEP}\ }\textbf {\bibinfo {volume} {01}},\ \bibinfo {pages} {107} (\bibinfo {year} {2017})},\ \Eprint {http://arxiv.org/abs/1611.03864} {arXiv:1611.03864 [hep-ph]} \BibitemShut {NoStop}%
\bibitem [{\citenamefont {Sung}\ \emph {et~al.}(2019)\citenamefont {Sung}, \citenamefont {Tu},\ and\ \citenamefont {Wu}}]{Sung:2019xie}%
  \BibitemOpen
  \bibfield  {author} {\bibinfo {author} {\bibfnamefont {A.}~\bibnamefont {Sung}}, \bibinfo {author} {\bibfnamefont {H.}~\bibnamefont {Tu}}, \ and\ \bibinfo {author} {\bibfnamefont {M.-R.}\ \bibnamefont {Wu}},\ }\href {\doibase 10.1103/PhysRevD.99.121305} {\bibfield  {journal} {\bibinfo  {journal} {Phys. Rev. D}\ }\textbf {\bibinfo {volume} {99}},\ \bibinfo {pages} {121305} (\bibinfo {year} {2019})},\ \Eprint {http://arxiv.org/abs/1903.07923} {arXiv:1903.07923 [hep-ph]} \BibitemShut {NoStop}%
\bibitem [{\citenamefont {DeRocco}\ \emph {et~al.}(2019)\citenamefont {DeRocco}, \citenamefont {Graham}, \citenamefont {Kasen}, \citenamefont {Marques-Tavares},\ and\ \citenamefont {Rajendran}}]{DeRocco:2019njg}%
  \BibitemOpen
  \bibfield  {author} {\bibinfo {author} {\bibfnamefont {W.}~\bibnamefont {DeRocco}}, \bibinfo {author} {\bibfnamefont {P.~W.}\ \bibnamefont {Graham}}, \bibinfo {author} {\bibfnamefont {D.}~\bibnamefont {Kasen}}, \bibinfo {author} {\bibfnamefont {G.}~\bibnamefont {Marques-Tavares}}, \ and\ \bibinfo {author} {\bibfnamefont {S.}~\bibnamefont {Rajendran}},\ }\href {\doibase 10.1007/JHEP02(2019)171} {\bibfield  {journal} {\bibinfo  {journal} {JHEP}\ }\textbf {\bibinfo {volume} {02}},\ \bibinfo {pages} {171} (\bibinfo {year} {2019})},\ \Eprint {http://arxiv.org/abs/1901.08596} {arXiv:1901.08596 [hep-ph]} \BibitemShut {NoStop}%
\bibitem [{\citenamefont {Bar}\ \emph {et~al.}(2020)\citenamefont {Bar}, \citenamefont {Blum},\ and\ \citenamefont {D'Amico}}]{Bar:2019ifz}%
  \BibitemOpen
  \bibfield  {author} {\bibinfo {author} {\bibfnamefont {N.}~\bibnamefont {Bar}}, \bibinfo {author} {\bibfnamefont {K.}~\bibnamefont {Blum}}, \ and\ \bibinfo {author} {\bibfnamefont {G.}~\bibnamefont {D'Amico}},\ }\href {\doibase 10.1103/PhysRevD.101.123025} {\bibfield  {journal} {\bibinfo  {journal} {Phys. Rev. D}\ }\textbf {\bibinfo {volume} {101}},\ \bibinfo {pages} {123025} (\bibinfo {year} {2020})},\ \Eprint {http://arxiv.org/abs/1907.05020} {arXiv:1907.05020 [hep-ph]} \BibitemShut {NoStop}%
\bibitem [{\citenamefont {Sung}\ \emph {et~al.}(2021)\citenamefont {Sung}, \citenamefont {Guo},\ and\ \citenamefont {Wu}}]{Sung:2021swd}%
  \BibitemOpen
  \bibfield  {author} {\bibinfo {author} {\bibfnamefont {A.}~\bibnamefont {Sung}}, \bibinfo {author} {\bibfnamefont {G.}~\bibnamefont {Guo}}, \ and\ \bibinfo {author} {\bibfnamefont {M.-R.}\ \bibnamefont {Wu}},\ }\href {\doibase 10.1103/PhysRevD.103.103005} {\bibfield  {journal} {\bibinfo  {journal} {Phys. Rev. D}\ }\textbf {\bibinfo {volume} {103}},\ \bibinfo {pages} {103005} (\bibinfo {year} {2021})},\ \Eprint {http://arxiv.org/abs/2102.04601} {arXiv:2102.04601 [hep-ph]} \BibitemShut {NoStop}%
\bibitem [{\citenamefont {Gupta}\ and\ \citenamefont {Sen}(2025)}]{Gupta:2025ygk}%
  \BibitemOpen
  \bibfield  {author} {\bibinfo {author} {\bibfnamefont {A.}~\bibnamefont {Gupta}}\ and\ \bibinfo {author} {\bibfnamefont {M.}~\bibnamefont {Sen}},\ }\href@noop {} {\  (\bibinfo {year} {2025})},\ \Eprint {http://arxiv.org/abs/2511.11219} {arXiv:2511.11219 [hep-ph]} \BibitemShut {NoStop}%
\bibitem [{\citenamefont {Krasny}(2015)}]{Krasny:2015ffb}%
  \BibitemOpen
  \bibfield  {author} {\bibinfo {author} {\bibfnamefont {M.~W.}\ \bibnamefont {Krasny}},\ }\href@noop {} {\  (\bibinfo {year} {2015})},\ \Eprint {http://arxiv.org/abs/1511.07794} {arXiv:1511.07794 [hep-ex]} \BibitemShut {NoStop}%
\bibitem [{\citenamefont {Krasny}\ \emph {et~al.}(2019)\citenamefont {Krasny}, \citenamefont {Martens},\ and\ \citenamefont {Dutheil}}]{Krasny:2019wch}%
  \BibitemOpen
  \bibfield  {author} {\bibinfo {author} {\bibfnamefont {M.~W.}\ \bibnamefont {Krasny}}, \bibinfo {author} {\bibfnamefont {A.}~\bibnamefont {Martens}}, \ and\ \bibinfo {author} {\bibfnamefont {Y.}~\bibnamefont {Dutheil}} (\bibinfo {collaboration} {Gamma Factory Study Group}),\ }\href@noop {} {\  (\bibinfo {year} {2019})}\BibitemShut {NoStop}%
\bibitem [{\citenamefont {Chung}(2024{\natexlab{c}})}]{Chung:2023iwj}%
  \BibitemOpen
  \bibfield  {author} {\bibinfo {author} {\bibfnamefont {Y.}~\bibnamefont {Chung}},\ }\href {\doibase 10.1103/PhysRevD.109.095021} {\bibfield  {journal} {\bibinfo  {journal} {Phys. Rev. D}\ }\textbf {\bibinfo {volume} {109}},\ \bibinfo {pages} {095021} (\bibinfo {year} {2024}{\natexlab{c}})},\ \Eprint {http://arxiv.org/abs/2309.00072} {arXiv:2309.00072 [hep-ph]} \BibitemShut {NoStop}%
\end{thebibliography}%

\end{document}